\documentclass{PoS}

\usepackage{epsfig,amsfonts,amsthm}
\usepackage[normalem]{ulem}
\usepackage{amsmath,amssymb}
\usepackage{array}
\usepackage{amsmath}
\usepackage{amsfonts}
\usepackage{amssymb}
\usepackage{subfig}
\usepackage{wrapfig}
\usepackage{graphicx}
\usepackage{slashed}
\usepackage{soul}
\newcommand{\be}{\begin{equation}}
\newcommand{\ee}{\end{equation}}
\newcommand{\bea}{\begin{eqnarray}}
\newcommand{\eea}{\end{eqnarray}}

\renewcommand{\Re}{\mathrm{Re }}
\renewcommand{\Im}{\mathrm{Im }}
\newcommand{\doublet}[2]{ \left( \begin{array}{c}#1 \\ #2 \end{array}\right) }

\usepackage[normalem]{ulem}
    
\definecolor{grey}{cmyk}{0,0,0,0.75}
\definecolor{tangerine}{cmyk}{0,0.5,1,0}
\definecolor{darkgreen}{cmyk}{1,0,1,0.23} 
\definecolor{Red}{rgb}{1,0,0}
\definecolor{Blue}{rgb}{0,0,1}
\definecolor{Green}{rgb}{0,1,0}
\definecolor{Grey}{cmyk}{0,0,0,0.75}
\definecolor{Tangerine}{cmyk}{0,0.5,1,0}
\definecolor{Darkgreen}{cmyk}{1,0,1,0.23}
\definecolor{Cyan}{cmyk}{1,0,0,0}
\definecolor{Yellow}{cmyk}{0,0,1,0}

\def\lsim{\mathrel{\rlap{\lower4pt\hbox{\hskip1pt$\sim$}}
    \raise1pt\hbox{$<$}}}         
\def\gsim{\mathrel{\rlap{\lower4pt\hbox{\hskip1pt$\sim$}}
    \raise1pt\hbox{$>$}}}         

\def\beq{\begin{equation}}
\def\eeq{\end{equation}}
\def\bea{\begin{eqnarray}}
\def\eea{\end{eqnarray}}

\def\<{\left\langle}
\def\>{\right\rangle}

\newcommand{\GeV}{{\ensuremath\rm \,GeV}}

\usepackage[normalem]{ulem}

\def\lsim{\mathrel{\rlap{\lower4pt\hbox{\hskip1pt$\sim$}}
    \raise1pt\hbox{$<$}}}         
\def\gsim{\mathrel{\rlap{\lower4pt\hbox{\hskip1pt$\sim$}}
    \raise1pt\hbox{$>$}}}         

\def\beq{\begin{equation}}
\def\eeq{\end{equation}}
\def\bea{\begin{eqnarray}}
\def\eea{\end{eqnarray}}

\def\<{\left\langle}
\def\>{\right\rangle}

\newcommand{\gev}{\mathrm{\;GeV}} 
\newcommand{\bt}{\begin{tabular}}
\newcommand{\et}{\end{tabular}}

\title{Dark origins of matter-antimatter asymmetry}

\ShortTitle{Dark CP violation}

\author{\speaker{Venus Keus}\\
        Department of Physics and Helsinki Institute of Physics,
  	Gustaf Hallstromin katu 2, FIN-00014 University of Helsinki, Finland\\
        E-mail: \email{venus.keus@helsinki.fi}}

\abstract{In a non-minimal Higgs framework, we present a novel mechanism in which the CP violating dark particles only interact with the SM through the gauge bosons, primarily the $Z$ boson. Such $Z$-portal dark CP violation is realised in the regions of the parameter space where Higgs-mediated (co)annihilation processes are sub-dominant and have negligible contributions to the DM relic density. 
We show that such $Z$-portal CP violating DM can still thermalise and satisfy all experimental and observational bounds and discuss the implications of such phenomena for electroweak baryogenesis.}

\FullConference{Corfu Summer Institute 2019 "School and Workshops on Elementary Particle Physics and Gravity" (CORFU2019)\\
		31 August - 25 September 2019\\
		Corfu, Greece}

\begin{document}

\section{Introduction}
\label{intro-section}

The great success of the Standard Model (SM) of particle physics was achieved with the discovery of the long-awaited Higgs boson, its last missing particle, in 2012 at the Large Hadron Collider (LHC) \cite{Aad:2012tfa,Chatrchyan:2012ufa}, as this particle was predicted by the electroweak symmetry breaking (EWSB) mechanism in 1964 \cite{Higgs:1964pj,Englert:1964et}.
The SM has been extensively tested and is in great agreement with experiment. So far, the LHC has not detected any significant deviation from the SM, and the observed scalar is very SM-like \cite{Flechl:2019jnr,Aad:2019mbh}. 

Nonetheless, it is widely accepted that the SM is not the ultimate theory of nature due to its shortcomings in explaining several phenomena;
many astrophysical observations hint towards the existence of a stable (on cosmological time scales) Dark Matter (DM) particle which is cold (non-relativistic at the onset of galaxy formation), non-baryonic, neutral and weakly interacting \cite{Ade:2015xua}. The particle content of SM does not provide such a candidate.  
Moreover, the SM lacks a successful baryogenesis mechanism explaining the origin of the observed matter-antimatter asymmetry in the universe. 
One of the most promising baryogenesis scenarios is electroweak baryogenesis (EWBG) \cite{Morrissey:2012db}, which produces the baryon excess during the electroweak phase transition (EWPT). Although the SM contains all required ingredients for EWBG, it is unable to explain the observed baryon excess due to its insufficient amount of CP violation \cite{Gavela:1993ts,Huet:1994jb,Gavela:1994dt} and the lack of a first-order phase transition \cite{Kajantie:1995kf}.
Another shortfall of the SM is that it offers no explanation from underlying physics which could give rise to the fermion masses and mixing patterns; in the SM the fermion masses and mixings are given by Yukawa couplings which are simply parameters to be measured.
The indisputable experimental evidence for neutrino masses and oscillations also necessitates going beyond the SM (BSM) framework.

A common characteristic of BSM theories, is an extended scalar sector, suggesting that the observed scalar at the LHC is just one member of an extended scalar sector.  
The scalar sector in the SM is the experimentally-least-constrained sector, and could provide new sources of CP violation and a strong first order EWPT if extended.
Moreover, non-minimal Higgs sectors with discrete symmetries could naturally accommodate Weakly Interacting Massive Particles (WIMPs) \cite{Jungman:1995df,Bertone:2004pz,Bergstrom:2000pn} as viable DM candidates whose stability is ensured by the conservation of the discrete symmetry after EWSB. 
The EWSB  patterns determine the number of active (developing a vacuum expectation value (VEV)) and inert (without a VEV) multiplets in the model.

Simple one-singlet and one-doublet extensions of the SM, \textit{i.e.} the Higgs portal model \cite{Bertolami:2007wb}, the 2-Higgs doublet model (2HDM) (see e.g. \cite{Branco:2011iw,Englert:2011yb} and references therein) and the Inert Doublet Model (IDM) \cite{Deshpande:1977rw}, have been studied to an advanced level, even though these models by construction can only partly provide a solution to some of the SM drawbacks.
In these models, 
either the scalar potential is inevitably CP-conserving\footnote{The purely singlet scalar extension of the SM is CP-conserving regardless of an apparent phase in the potential or the vacuum \cite{Branco:1999fs}.} due to an exact $Z_2$ symmetry to stabilise the DM candidate, or CP violation is introduced at the expense of breaking the $Z_2$ symmetry and loosing the DM candidate as a result. Nevertheless, such new sources of CP-violation modify the SM-Higgs couplings and 
contribute to the Electric Dipole Moments (EDMs) of the neutron, electron, and certain atomic nuclei \cite{Chupp:2017rkp} and are, therefore, highly constrained by experiment \cite{Inoue:2014nva,Keus:2015hva,Keus:2017ioh,Yamanaka:2017mef}.

One needs to go beyond the simple doublet or singlet scalar extensions of the SM to incorporate both CP-violation and DM into the model. 
To avoid the pitfalls of the IDM and the Higgs portal model, one needs to extend the inert sector which contains the DM candidate whose stability is ensured by the symmetry of the potential, or remnant thereof after EWSB.
If one were to extend the active sector to accommodate CP-violation, one runs into the same limitations as in the 2HDM due to the modification of the SM-Higgs couplings and 
contributions to the EDMs  \cite{Grzadkowski:2009bt, Osland:2013sla}.
On the other hand, if CP violation is introduced in the extended inert sector, there will be no contributions to the EDMs and no limit on the amount of CP violation in the model, since the inert sector is protected from directly coupling to the SM particles.

The phenomena of \textit{dark CP-violation} and \textit{CP violating DM} were introduced for the first time in \cite{Cordero-Cid:2016krd,Keus:2016orl,Cordero:2017owj,Cordero-Cid:2018man,Keus:2019szx}, and was shown to 
provide a handle on the otherwise fixed gauge couplings. 
As a result, a large region of the parameter space opens up to accommodate a DM candidate in agreement with cosmological and collider experiments.
Note that a 3-Higgs doublet model (3HDM) with an extended inert sector is superior to a 2HDM plus a singlet scalar
with an extended inert sector \cite{Azevedo:2018fmj}; the amount of CP-violation is inevitably reduced due to the presence of the singlet and there exist fewer co-annihilation channels  for the DM candidate.

To build a stronger case for 3HDMs, let us point out that not only they 
contain viable DM candidates, open up a whole new possibility to the CP quantum number of the DM candidate, provide new sources of CP-violation and strongly first order EWPT \cite{Ahriche:2015mea}, but also provide a solution to the fermion mass hierarchy problem \cite{Howl:2009ds,Felipe:2013ie,Morisi:2011pt,Ding:2013hpa,King:2013hj}, all in one framework owing to different symmetries that can be imposed on the scalar potential \cite{Keus:2019szx}. 
Here, we expand on the dark CP violation mechanism, and require the dark/inert particles to interact with the SM only  through the gauge bosons, primarily the $Z$ boson.
The Higgs-DM couplings is heavily constrained by direct and indirect detection experiments and the LHC bound on the BR$(h \to inv.)$ for low DM masses, $m_{DM} < m_h/2$. We show that the $Z$ portal CP-violating DM which is rid of these constraints, can still thermalise and satisfy all experimental and observational data. 

In what follows, we present the details of the scalar potential and the theoretical and experimental limits on its parameters in section \ref{scalar-potential}. We construct and justify our benchmark scenarios in section \ref{selection}. The effect of dark CP-violation on the production and annihilation of DM is discussed in section \ref{Abundance}, and finally, we conclude and present the outlook for our future studies in section \ref{conclusion}.

\section{The scalar potential with explicit CP violation}
\label{scalar-potential}

A 3HDM scalar potential symmetric under a group $Z_2$ can be written as the sum of $V_0$ with terms symmetric under any phase rotation, and $V_{Z_2}$ with terms ensuring the symmetry of the potential \cite{Ivanov:2011ae,Keus:2013hya}, $V_{3HDM}=V_0+V_{Z_2}, $ where
\bea
V_0 &=& - \mu^2_{1} (\phi_1^\dagger \phi_1) -\mu^2_2 (\phi_2^\dagger \phi_2) - \mu^2_3(\phi_3^\dagger \phi_3)
+ \lambda_{11} (\phi_1^\dagger \phi_1)^2+ \lambda_{22} (\phi_2^\dagger \phi_2)^2  + \lambda_{33} (\phi_3^\dagger \phi_3)^2 
\nonumber\\
&& 
+ \lambda_{12}  (\phi_1^\dagger \phi_1)(\phi_2^\dagger \phi_2)
 + \lambda_{23}  (\phi_2^\dagger \phi_2)(\phi_3^\dagger \phi_3) + \lambda_{31} (\phi_3^\dagger \phi_3)(\phi_1^\dagger \phi_1)
 + \lambda'_{12} (\phi_1^\dagger \phi_2)(\phi_2^\dagger \phi_1) 
  \nonumber\\
&& 
 + \lambda'_{23} (\phi_2^\dagger \phi_3)(\phi_3^\dagger \phi_2) + \lambda'_{31} (\phi_3^\dagger \phi_1)(\phi_1^\dagger \phi_3),  \nonumber\\
 V_{Z_2} &=& -\mu^2_{12}(\phi_1^\dagger\phi_2)+  \lambda_{1}(\phi_1^\dagger\phi_2)^2 + \lambda_2(\phi_2^\dagger\phi_3)^2 + \lambda_3(\phi_3^\dagger\phi_1)^2  + h.c., 
 \label{V0-3HDM}
\eea
where the three Higgs doublets, $\phi_{1},\phi_2,\phi_3$, transform under the $Z_2$ group as: 
\be 
\label{generator}
\phi_1 \to - \phi_1, \quad 
\phi_2 \to - \phi_2, \quad
\phi_3 \to +\phi_3 \, .
\ee
The $Z_2$ symmetry is respected by the vacuum,
\be 
\langle \phi_1 \rangle =0, \quad  \langle \phi_2 \rangle =0, \quad 
\langle \phi_3 \rangle =\frac{v}{\sqrt{2}},
\label{vacuum}
\ee
rendering $\phi_1$ and $\phi_2$ as the $Z_2$-odd inert doublets and $\phi_3$ as the $Z_2$-even active doublet.
The composition of the doublets is as follows,
\be 
\phi_1= \doublet{$\begin{scriptsize}$ H^+_1 $\end{scriptsize}$}{\frac{H_1+iA_1}{\sqrt{2}}},\quad 
\phi_2= \doublet{$\begin{scriptsize}$ H^+_2 $\end{scriptsize}$}{\frac{H_2+iA_2}{\sqrt{2}}}, \quad 
\phi_3= \doublet{$\begin{scriptsize}$ G^+ $\end{scriptsize}$}{\frac{v+h+iG^0}{\sqrt{2}}}, 
\label{explicit-fields}
\ee
where $\phi_3$ plays the role of the SM Higgs doublet, with $h$ being the SM Higgs boson and $G^\pm,~ G^0$ the would-be Goldstone bosons. 

The parameters of the phase invariant part of the potential, $V_0$, are by construction real, while the parameters of  $V_{Z_2}$, namely $\mu^2_{12}, \lambda_1,\lambda_2, \lambda_3$, could be complex and act as a source of explicit CP violation\footnote{Note that $\lambda_1$ (and other dark sector parameters $\lambda_{11},\lambda_{22},\lambda_{12}, \lambda'_{12}$)  appear only in the inert scalars self-interactions and have no relevance for our DM and collider phenomenology studies. Therefore, for simplicity, their values are set to $0.1$ which respects the bounds from perturbative unitarity and positivity of the potential.}.
Thus, the CP-violation is introduced in the inert/dark sector which is forbidden, by the conservation of the $Z_2$ symmetry, from mixing with the  active sector, and is therefore not limited by EDMs. The lightest particle amongst the CP-mixed neutral fields from the inert doublets is a stable particle and a viable DM candidate. 
The phenomenologically relevant parameters are $\mu^2_3, \lambda_{33}$ which are fixed by the Higgs mass, and $\mu^2_{1},\mu^2_{2},\mu^2_{12}, \lambda_{31},\lambda_{23},\lambda'_{31},\lambda'_{23},\lambda_{2}, \lambda_{3}$ which appear in inert scalars masses and couplings and are, in principle, independent. 
However, here we limit our study to the \textit{dark democracy} limit \cite{Keus:2014jha,Keus:2015xya,Cordero-Cid:2016krd, Cordero-Cid:2018man}, where
\be 
\mu^2_1 =\mu^2_2 , \quad \lambda_3=\lambda_2 , \quad \lambda_{31}=\lambda_{23} ,\quad \lambda'_{31}=\lambda'_{23}.
\ee
Note that after imposing the dark democracy limit, the model is still explicitly CP-violating since $(\lambda_{22}- \lambda_{11} )
\left[\lambda_1 ({\mu^2_{12}}^*)^2-\lambda^*_{1}(\mu^2_{12})^2 \right] \neq 0$ \cite{Haber:2006ue,Haber:2015pua}. 
Further, one could rotate away the apparent phase of $\mu^2_{12}$ by a redefinition of doublets \cite{Cordero-Cid:2018man}. This leaves $\theta_2$, the phase of the $\lambda_2$ parameter, 
\be  
\label{notation}
\lambda_2 = \Re \lambda_2 +i  \Im\lambda_2 = |\lambda_2| e^{i \theta_2},
\ee
as the only relevant CP-violating factor which is referred to as $\theta_{\rm CPV}$ throughout the paper.

\subsection{The mass spectrum}
\label{minimization}

The minimum of the potential lies at the $(0,0,\frac{v}{\sqrt{2}})$ point when $ v^2= \mu^2_3/\lambda_{33}$.
The fields in the only active doublet, $\phi_3$, are mass eigenstates with $G^0,G^\pm$ as the massless Goldstone bosons, and $h$ as the SM-like Higgs with
$ m^2_{h}= 2\mu_3^2 =2\lambda_{33} v^2 = (125$ GeV$)^2$.

The inert doublets mix, resulting in two physical charged states, $S^\pm_{1,2}$, which are a combination of the charged components of the inert doublets, $H^\pm_{1,2}$, 
\be 
S^\pm_{1}= \frac{H^\pm_{1} + H^\pm_{2}}{\sqrt{2}}, \qquad
S^\pm_{2}= \frac{H^\pm_{1} - H^\pm_{2}}{\sqrt{2}},
\ee
with mass-squared values
\be 
m^2_{S^\pm_{1}}
=
 - \mu_2^2 - \mu_{12}^2 +\frac{1}{2}\lambda_{23}v^2, \qquad
m^2_{S^\pm_{2}}
=
- \mu_2^2 + \mu_{12}^2 + \frac{1}{2}\lambda_{23}v^2 ,
\ee
where require $\mu_{12}^2 >0$ and fix the hierarchy of the inert charged states, $m_{S_1^\pm} < m_{S_2^\pm}$. 
The four inert neutral scalars, $S_{1},S_{2}, S_{3},S_{4}$, are a combination of the CP-even and CP-odd components of the inert doublets, $H_1, H_2,A_1, A_2$, 
\bea 
&& 
S_1 =\frac{\alpha H_1 -A_1+ \alpha H_2+A_2}{\sqrt{2} \; \sqrt{\alpha^2+1}},\qquad
S_2 =\frac{H_1 +\alpha A_1+ H_2  - \alpha A_2}{\sqrt{2} \; \sqrt{\alpha^2+1}},  \\[2mm]
&& 
S_3 =\frac{\beta H_1+A_1 -\beta H_2+A_2}{\sqrt{2} \; \sqrt{\beta^2+1}},
\qquad
S_4 =\frac{- H_1+\beta A_1  + H_2 +\beta A_2}{\sqrt{2}\; \sqrt{\beta^2+1}}, \nonumber
\eea
where $\alpha$ and $\beta$ are defined as
\be
\label{alpha-beta} 
\alpha 
= 
\frac{
- \mu^2_{12}
+v^2 | \lambda_2|  \cos\theta_{CPV}
-\Lambda^- }{v^2 | \lambda_2|  \sin\theta_{CPV}}
,\qquad
\beta = \frac{
-\mu^2_{12}
-v^2 | \lambda_2|  \cos\theta_{CPV}
+ \Lambda^+ }{v^2 | \lambda_2|  \sin\theta_{CPV}},
\ee
and $\Lambda^{\mp} $ as
\be  
\label{lambdas}
\Lambda^{\mp} =\sqrt{(\mu^2_{12})^2 + v^4|\lambda_2|^2 \mp 2 v^2 \mu^2_{12} |\lambda_2|\cos\theta_{CPV}} \; .
\ee
The masses of the neutral CP-mixed inert scalars, $S_{1},S_{2}, S_{3},S_{4}$, are calculated to be
\be 
\label{masses-Ss}
m^2_{S_{1,2}} =  -\mu^2_2
+\frac{v^2}{2}(\lambda'_{23}+\lambda_{23}) \mp \Lambda^- ,
\qquad
m^2_{S_{3,4}} =  -\mu^2_2
+\frac{v^2}{2}(\lambda'_{23}+\lambda_{23}) \mp \Lambda^+  .
\ee
As the DM candidate, we require $S_1$ to be lightest inert particle which leads to 
\be 
m_{S_1} < m_{S_2}, m_{S_3},m_{S_4} \qquad \Rightarrow \qquad
\frac{\pi}{2} < \theta_{CPV} < \frac{3\pi}{2},
\ee
when $\lambda_2 < 0$ in agreement with \cite{Keus:2014jha,Keus:2015xya}
\footnote{If $\theta_{CPV}$ is taken to be in the 2$^{nd}$ quadrant, for $\lambda_2 > 0$ the above augments hold provided the neutral inert particles are relabelled as $S_1 \leftrightarrow S_3$ and $S_2 \leftrightarrow S_4$, similarly, if $ \theta_{CPV}$ is taken to be in the 1$^{st}$ or 4$^{th}$ quadrants for $\lambda_2 < 0$.}. 
At $\theta_{CPV} = \frac{\pi}{2}, \frac{3\pi}{2}$ where $\Lambda^+=\Lambda^-$, a mass degeneracy between neutral inert particles occurs, $m_{S_1} = m_{S_3}$ and $m_{S_2} =m_{S_4}$.
At $\theta_{CPV} = 0,\pi$, the model is reduced to the CP-conserving limit which renders $S_{1,3}$ to CP-even and $S_{2,4}$ to CP-odd particles,
\be 
\theta_{CPV} = 0,\pi \qquad \Rightarrow \qquad
\mbox{CP-conserving limit}: \;
\left\{ \begin{array}{c}
S_{1,3} = \frac{H_1 \pm H_2}{\sqrt{2}},\\[2mm]
S_{2,4} =\frac{A_1 \pm A_2}{\sqrt{2}}.\\
\end{array}
\right. 
\label{CPC-limit}
\ee
We take all other parameters of the potential to be positive. 
The independent parameters, $\mu^2_{12}$, $\lambda_{23}$, $\lambda'_{23}$, $\mu^2_2$, $\lambda_2$,
are written in terms of scalar masses and couplings, $m_{S_1}$, $m_{S_2}$, $m_{S^\pm_{1}}$, $m_{S^\pm_{2}}$, $g_{hDM}$, 
which are taken as the input parameters of the model, where $g_{hDM} \equiv g_{S_1S_1h}$ is the Higgs-DM coupling, with the relevant terms in the Lagrangian appearing as:
\be 
\mathcal{L} \; \supset \; g_{ZS_iS_j} Z_\mu (S_i \partial^\mu S_j - S_j \partial^\mu S_i) +  \;
\frac{v}{2}g_{S_i S_i h} h S_i^2+ \;
v g_{S_i S_j h} h S_i S_j +  \;
v g_{S_i^\pm S_j^\mp h} h S_i^\pm S_j^\mp. \label{ghSS}
\ee

\subsection{Constraints on the parameter space}
\label{constraints}

The parameter space of the model is constrained by: 
\begin{enumerate}
\item 
theoretical bounds: 
boundedness-from-below of the potential, positive-definiteness of the Hessian, perturbative unitarity and electroweak oblique $S,T,U$ parameters
\item 
experimental bounds:
total decay width of $W^\pm,Z$ bosons, non-observation of charged scalars, Higgs total decay width and Higgs signal strengths, Higgs invisible branching ratio for light inert scalars, 
\item 
observational bounds: relic density measurements and direct and indirect detection of DM,
\end{enumerate}
which are satisfied in all our benchmark scenarios as detailed in \cite{Keus:2019szx}.

\section{The abundance of DM and benchmark selection}
\label{selection}

The solution of the Boltzmann equation after freeze out, determines the relic abundance of the DM candidate, $S_1$:
\be 
\frac{d n_{S_1}}{dt} = 
- 3\, H \,n_{S_1} - \langle \sigma_{eff}\, v \rangle \, \left[ (n_{S_1})^2 - (n^{eq}_{S_1})^{2} \right],
\ee
where $n_{S_1}$ ($n^{eq}_{S_1}$) is the number density of the $S_1$ particle (at equilibrium), and $H$ is the Hubble parameter. 
The thermally averaged cross section, $\langle \sigma_{eff}\, v \rangle$, receives contribution from all relevant (co)annihilation processes of any $S_i S_j$ pair into SM particles, so that
\be 
\langle \sigma_{eff} v \rangle = 
\sum_{i,j} \langle \sigma_{ij}\, v_{ij} \rangle \,\frac{n^{eq}_{S_i}}{n^{eq}_{S_1}} \, \frac{n^{eq}_{S_j}}{n^{eq}_{S_1}},
\qquad
\mbox{where}
\qquad
\frac{n^{eq}_{S_i}}{n^{eq}_{S_1}} \sim \exp({-\frac{m_{S_i} - m_{S_1}}{T}}),
\ee
with the main contribution from processes where $m_{S_i} - m_{S_1}$ is comparable to the thermal bath temperature $T$.

A common feature of non-minimal Higgs DM models is that in a large region of the parameter space the dominant process for DM annihilation is
$S_1 S_1 \to h_{\rm SM} \to f \bar f$
whose efficiency depends both on the DM mass and the Higgs-DM coupling. 
In the low mass region, $m_{\rm DM} < m_h/2$, generally one requires a large Higgs-DM coupling for an effective DM annihilation leading to a relic density in agreement with observations.
However, such large Higgs-DM couplings result in large direct and indirect detection cross sections and significant deviations from SM-Higgs coupling measurements, and hence are ruled out by experimental and observational data. 
On the other hand, a small Higgs-DM coupling, fails to annihilate DM effectively and results in the over-closure of the universe.
Here, the co-annihilation processes play an important role as they can contribute to changes in the relic density of DM.
 
In models with extended dark sectors, in addition to the standard Higgs mediated DM annihilation channels, there exist such co-annihilation channels of DM with heavier states, provided they are close in mass \cite{Cordero-Cid:2016krd,Cordero:2017owj,Cordero-Cid:2018man,Keus:2014jha,Keus:2015xya}. The relevance of these processes depends not only on the DM mass and the mass splittings, but also on the strength of the standard DM annihilation process. 
It is worth emphasising that in the IDM, which is by construction CP conserving, the only possible co-annihilation process is through the $Z$-mediated $H\,A \to Z \to f \bar f$ channel whose sub-dominant effect fails to rescue the model in the low mass region where it is in severe tension with direct and indirect detection bounds and BR($h \to inv.$) limits.

Extending the inert sector,
as shown in \cite{Cordero:2017owj,Keus:2014jha,Keus:2015xya} already in the CP-conserving limit, allows for several co-annihilation channels, both Higgs-mediated $H_1\,H_2 \to h \to f \bar f$ and $Z$-mediated $H_1\,A_{1,2} \to Z \to f \bar f$. The collective contribution of these channels to DM co-annihilation is not sufficient to completely rid the model of experimental and observational constraints, since one still needs a non-zero Higgs-DM coupling to satisfy relic density bounds.
Introducing CP-violation in the extended dark sector \cite{Cordero-Cid:2016krd,Cordero-Cid:2018man,Fuyuto:2019vfe}, triumphantly opens up many co-annihilation channels through the Higgs and $Z$ bosons, $S_i\,S_j \to h/Z \to f \bar f$, which can significantly affect the DM phenomenology.
In fact, the $Z$-mediated co-annihilations can be strong enough to relieve the model of the need for any Higgs-mediated (co)annihilation processes \cite{Keus:2019szx}. 

Focusing on the regions of the parameter space where the Higgs-DM coupling is negligibly small, $g_{hDM}\sim 0$, we highlight the effect of $Z$ portal CP-violation on the abundance of DM.
Here, the main DM annihilation process, $S_1 S_1 \to h \to f \bar f$ and, as a result, the Higgs-mediated co-annihilation processes are sub-dominant and have negligible contributions to the DM relic density. 
Therefore, the only effective communication between the dark sector and the visible sector is through the gauge bosons $W^\pm$ and $Z$ .

Let us emphasise that the phenomenon of dark CP violation, is not realisable in purely scalar singlet extensions of the SM. 
An extended dark sector with a doublet plus a singlet could, in principle, accommodate dark CP-violation, however, the presence of the singlet dilutes the CP violating effects, since a singlet has no direct coupling to SM gauge bosons. 
As a result, the model fails to provide a DM candidate accounting for 100\% of the observed relic density in the low mass region \cite{Azevedo:2018fmj}
We would like to point out that this is the reason Ref. \cite{Azevedo:2018fmj} fails to find a DM candidate accounting for 100\% of the observed relic density in the low mass region. Furthermore, the collider signatures of dark CP violation through the $ZZZ$ vertex and cross section asymmetries \cite{Cordero-Cid:2018man,Grzadkowski:2016lpv,Cordero-Cid:2020yba}, in their model is considerably smaller.

With the negligible Higgs mediated processes, the total DM annihilation cross section is determined by the following gauge boson mediated processes, $V=Z, W^\pm$:
\begin{itemize}
\item 
\textbf{DM annihilation processes:}
\be 
S_1 S_1 \to V V , \qquad
S_1 S_1 \to V V^* \to V f f', \qquad 
S_1 S_1 \to V^* V^* \to f f' f f' \,,
\label{annihilation-1}
\ee
where the processes with off-shell gauge bosons dominate over the ones with on-shell gauge bosons, in the $m_{DM} < m_W$ region.

\item 
\textbf{DM co-annihilation processes:}
\be 
S_1 S_{2,3,4} \to Z^* \to f \bar f, \qquad S_1 S^\pm_{1,2} \to W^{\pm *} \to f f' \, ,
\label{annihilation-2}
\ee
where the co-annihilating dark scalars are up to 20\% heavier than the DM candidate.

\item 
\textbf{(co)annihilation of other dark states:}
\be  
S_i S_i \to V V , \quad
S_i S_i \to V V^* \to V f f', \quad 
S_i S_i \to V^* V^* \to f f' f f', \quad 
S_i S_j \to V^* \to f f', 
\label{annihilation-3}
\ee
where $S_i\neq S_j$ are any of the dark scalars $ S_{2,3,4}\,, S^\pm_{1,2}$ which are all close in mass.
\end{itemize}

Taking all the above processes into account, we define the following characteristic benchmark scenarios with distinct DM phenomenology.
We introduce the notations
\be
\delta_{12} = m_{S_2} - m_{S_1}, \qquad \delta_{c} = m_{S_2^\pm} - m_{S_1^\pm} , \qquad \delta_{1c} = m_{S_1^\pm} - m_{S_1}\, ,
\ee
representing the mass splittings between the DM candidate and other inert scalars.

\subsubsection*{Benchmarks of type 1:}
\noindent
In the low mass region, $45 \gev <m_{S_1} \leq 80 \gev$, we devise benchmark scenarios of type 1:
\bea
\mbox{\textbf{B$_1$D$_{4}$C$_{1}$}} &:& \; \delta_{12} =4 \GeV,\quad \delta_{c} = 1\GeV,\quad \delta_{1c} = 50\GeV ,  \nonumber \\
\mbox{\textbf{B$_1$D$_{8}$C$_{1}$}} &:& \; \delta_{12} =8 \GeV,\quad \delta_{c} = 1\GeV,\quad \delta_{1c} = 50\GeV , \nonumber  \\
 \mbox{where}~&&~  m_{S_1} \sim m_{S_3}\sim m_{S_2} \sim m_{S_4} \ll m_{S_1^\pm} \sim m_{S_2^\pm} \, ,
\eea
with all neutral inert particles close in mass, and much lighter than the inert charged particles. 
We define another two benchmark scenarios of type 1 with a larger $\delta_{12}$, where the neutral inert particles split into two groups, with $S_1$ and $S_3$ close in mass and lighter than $S_2$ and $S_4$ which are also close in mass, and with all neutral inert scalars lighter than the charged inert scalars,
\bea
\mbox{\textbf{B$_1$D$_{12}$C$_1$}} &:& \; \delta_{12} =12 \GeV, \quad\delta_{c} = 1\GeV, \quad\delta_{1c} = 50\GeV,  
\nonumber  \\
\mbox{\textbf{B$_1$D$_{20}$C$_1$}} &:& \; \delta_{12} =20 \GeV, \quad\delta_{c} = 1\GeV, \quad\delta_{1c} = 50\GeV\, \nonumber\\
 \mbox{where}~&&~  
 m_{S_1} \sim m_{S_3} \; \lesssim \;  m_{S_2} \sim m_{S_4} \;\ll \; m_{S_1^\pm} \sim m_{S_2^\pm}  \,.
\eea

\subsubsection*{Benchmarks of type 2:}
\noindent
In the low mass region, $45 \gev <m_{S_1} \leq 80 \gev$, we devise two benchmark scenarios of type 2:
\bea
\mbox{\textbf{B$_2$D$_{55}$C$_{1}$}} &:& \; \delta_{12} =55 \GeV, \quad \delta_{c} = 1\GeV, \quad \delta_{1c} = 50\GeV, \nonumber  \\
\mbox{\textbf{B$_2$D$_{55}$C$_{15}$}} &:& \; \delta_{12} =55 \GeV, \quad \delta_{c} = 15\GeV, \quad \delta_{1c} = 50\GeV,  \nonumber \\
 \mbox{where}~&&~  
m_{S_1} \sim m_{S_3} \ll m_{S_2} \sim m_{S_4} \sim  m_{S_1^\pm} \sim m_{S_2^\pm} \, 
\eea
where only one neutral inert particle, $S_3$, is close in mass with the DM candidate, $S_1$.

\subsubsection*{Benchmarks of type 3:}
\noindent
In the heavy mass region $m_{S_1} \geq 80 \gev$, where the DM candidate can be close in mass with the charged inert particles, we define three benchmark scenarios
\bea
\mbox{\textbf{B$_3$D$_{5}$C$_1$}} 
&:&
\; \delta_{12} =5 \GeV,\quad \delta_{c} = 1\GeV, \quad \delta_{1c} = 1\GeV, 
\nonumber\\
\mbox{where} ~ && ~  
m_{S_1} \sim m_{S_3} \sim m_{S_2} \sim m_{S_4} \sim  m_{S_1^\pm} \sim m_{S_2^\pm} \,, 
\\
\mbox{\textbf{B$_3$D$_{55}$C$_1$} }
&:&
\; \delta_{12} =55 \GeV, \quad\delta_{c} = 1\GeV, \quad\delta_{1c} = 1\GeV,  
\nonumber \\
\mbox{where}~&&~  
m_{S_1} \sim m_{S_3} \sim  m_{S_1^\pm} \sim m_{S_2^\pm} \ll m_{S_2} \sim m_{S_4} \,,
\\
\mbox{\textbf{B$_3$D$_{55}$C$_{22}$}} 
&:&
\; \delta_{12} =55 \GeV, \quad\delta_{c} = 22 \GeV, \quad\delta_{1c} = 1 \GeV\, ,
\nonumber\\
\mbox{where}~&&~   
m_{S_1} \sim m_{S_3} \sim  m_{S_1^\pm}  \ll  m_{S_2^\pm} \sim m_{S_2} \sim m_{S_4} \,.
\eea

\section{The effect of dark CP-violation on the DM abundance}
\label{Abundance}
\subsection{Benchmarks of type 1}
\begin{figure}[h!]
\centering
\includegraphics[scale=0.47]{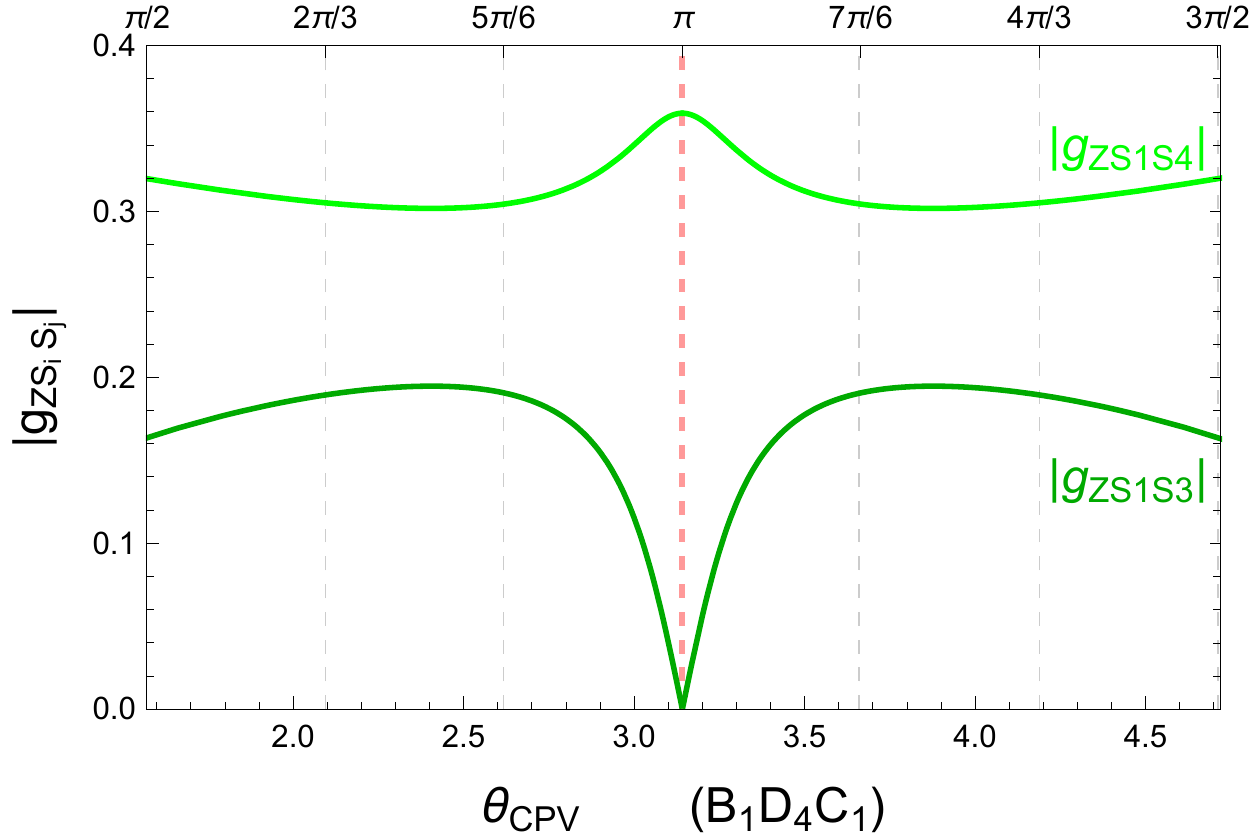}
\hspace{1mm}
\includegraphics[scale=0.47]{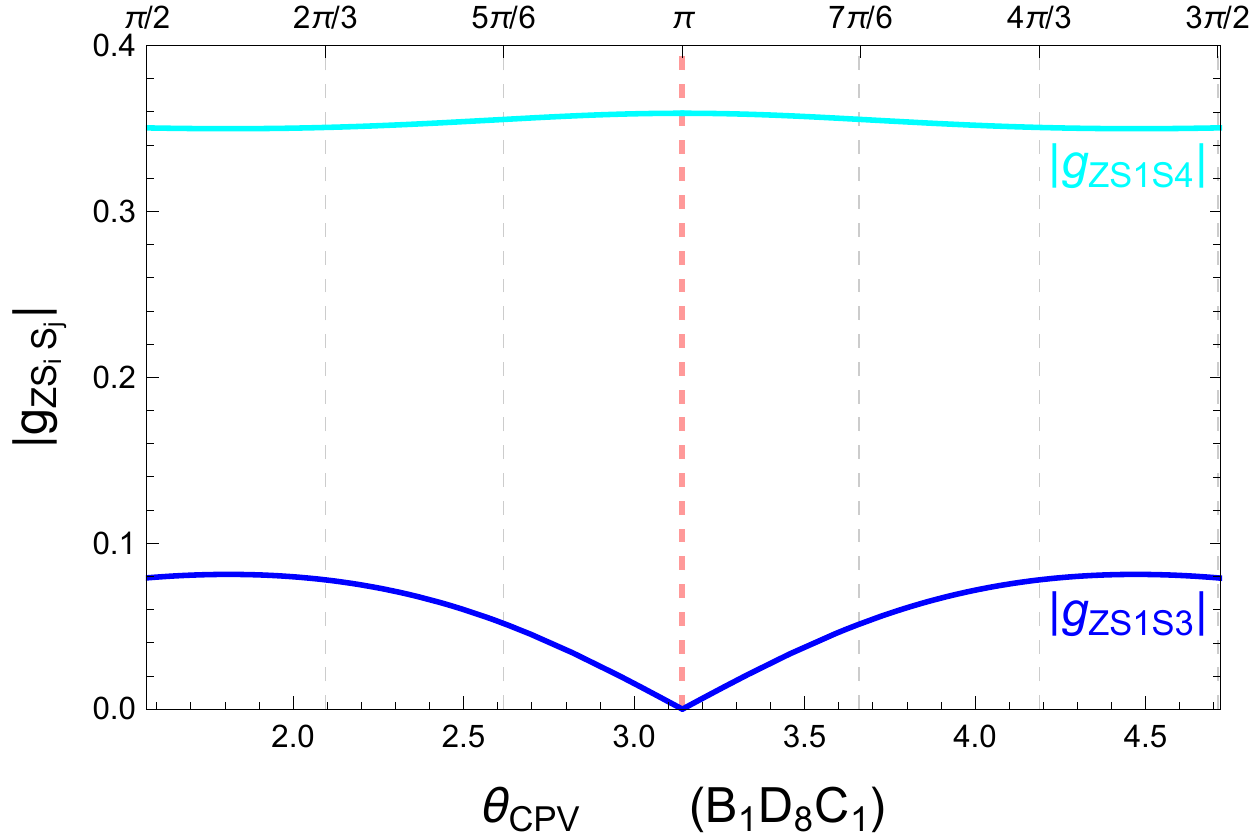}\\[2mm]
\includegraphics[scale=0.47]{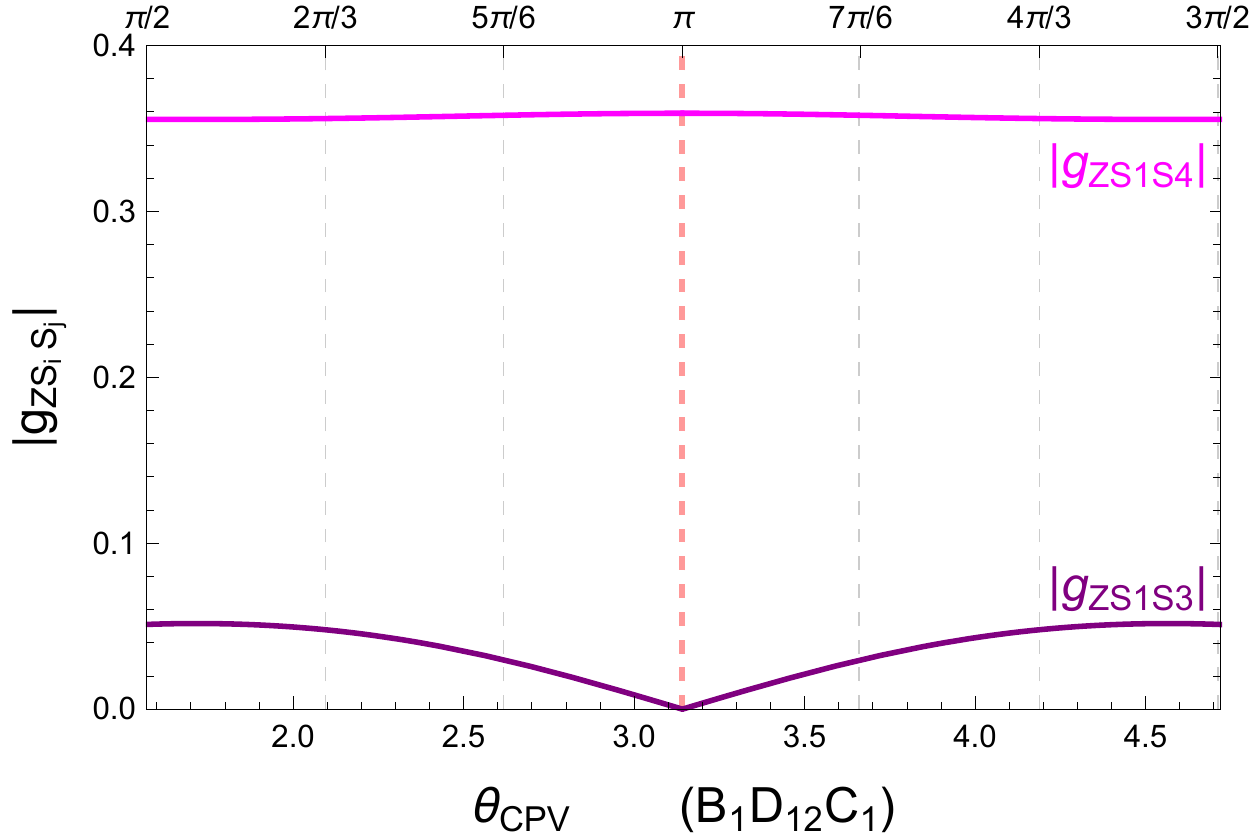}
\hspace{1mm}
\includegraphics[scale=0.47]{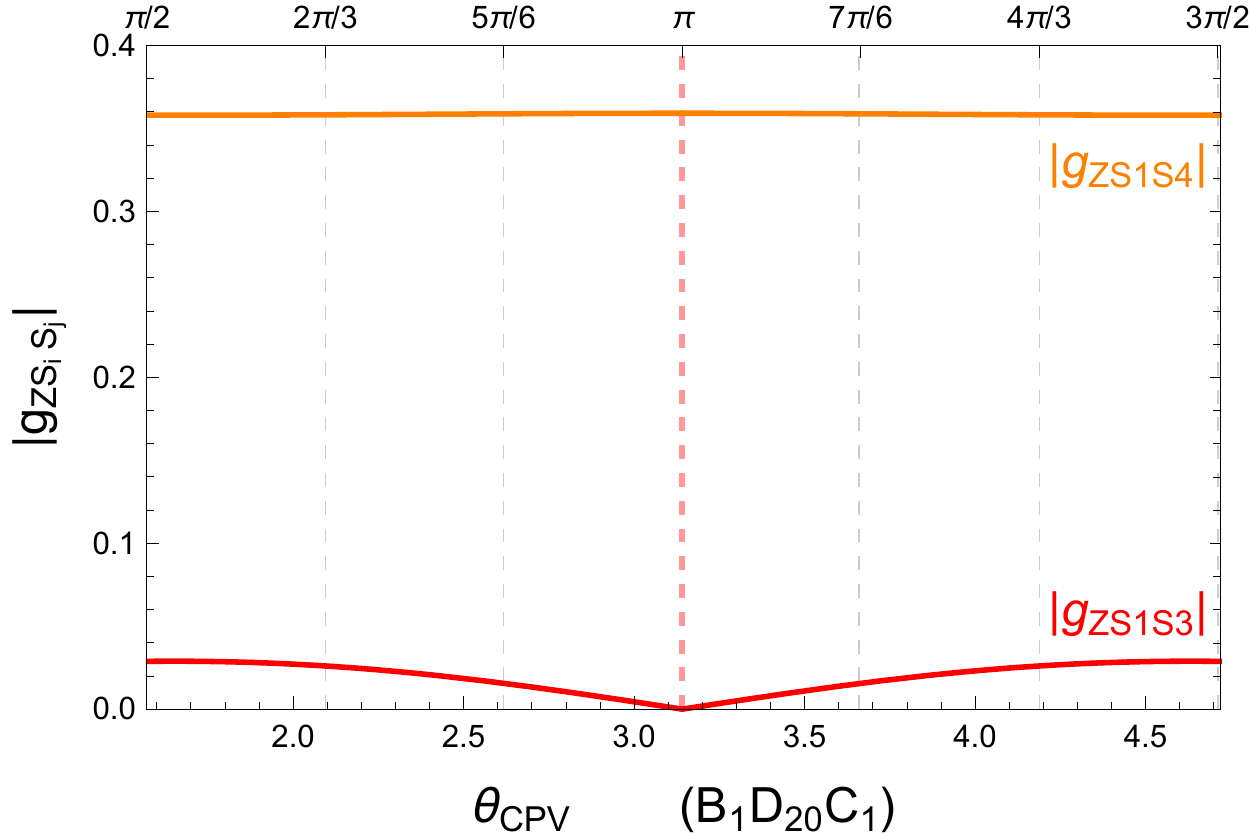}
\caption{The relevant $ZS_iS_j$ couplings in type 1 scenarios for an exemplary $m_{S_1}=57$ GeV.}
\label{B1-gZSiSj-fig}
\end{figure}

Benchmark scenarios of type 1, represent regions of the parameter space where all neutral inert particles are relatively close in mass and are much lighter than the inert charged particles.
Therefore, the main co-annihilation channel in these scenarios is through the
$S_iS_j \to Z^* \to f \bar f$ processes. 
In B$_1$D$_{4}$C$_{1}$ and B$_1$D$_{8}$C$_{1}$ scenarios, with very small $S_1-S_2$ mass splitting, the co-annihilation of DM with other neutral scalars is so strong that DM is under-produced irrespective of size of the CP violating angle, $\theta_{CPV}$.
In B$_1$D$_{12}$C$_{1}$ and B$_1$D$_{20}$C$_{1}$ scenarios, with a larger $S_1-S_2$ mass splitting, these co-annihilation processes are weakened and the $S_1$ relic abundance is increased as a result.
The efficiency of the co-annihilation processes is also dependant on the strength of the $ZS_iS_j$ coupling.
Figure \ref{B1-gZSiSj-fig} shows the strength of the relevant and non-negligible $ZS_iS_j$ couplings in all four type 1 scenarios for an exemplary $m_{S_1}$ of 57 GeV. 
As expected the $g_{ZS_1S_3}$ coupling vanishes at $\theta_{CPV}=\pi$ which is the CP-conserving limit where $S_1$ and $S_3$ are reduced to two CP-even particles as shown in eq.(\ref{CPC-limit}). 

\begin{figure}[h!]
\centering
\includegraphics[scale=0.47]{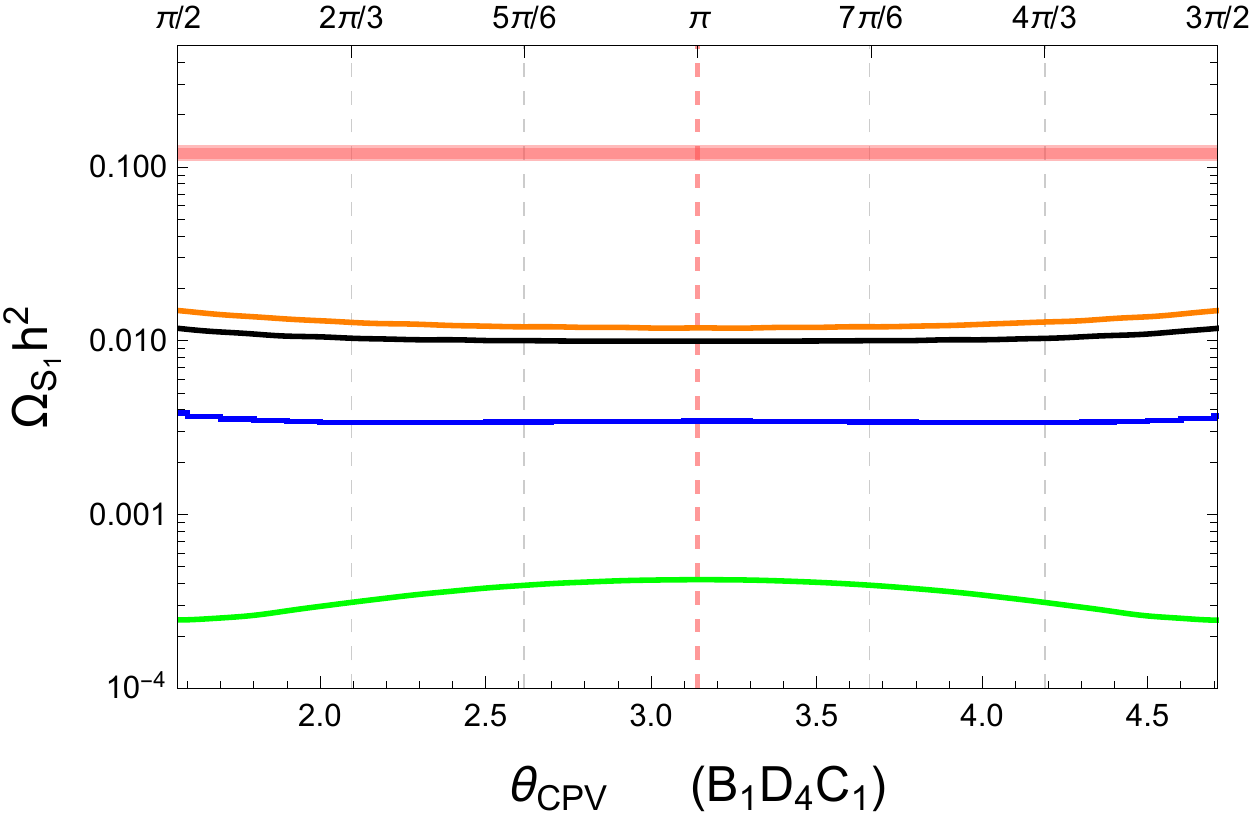}
\hspace{1mm}
\includegraphics[scale=0.47]{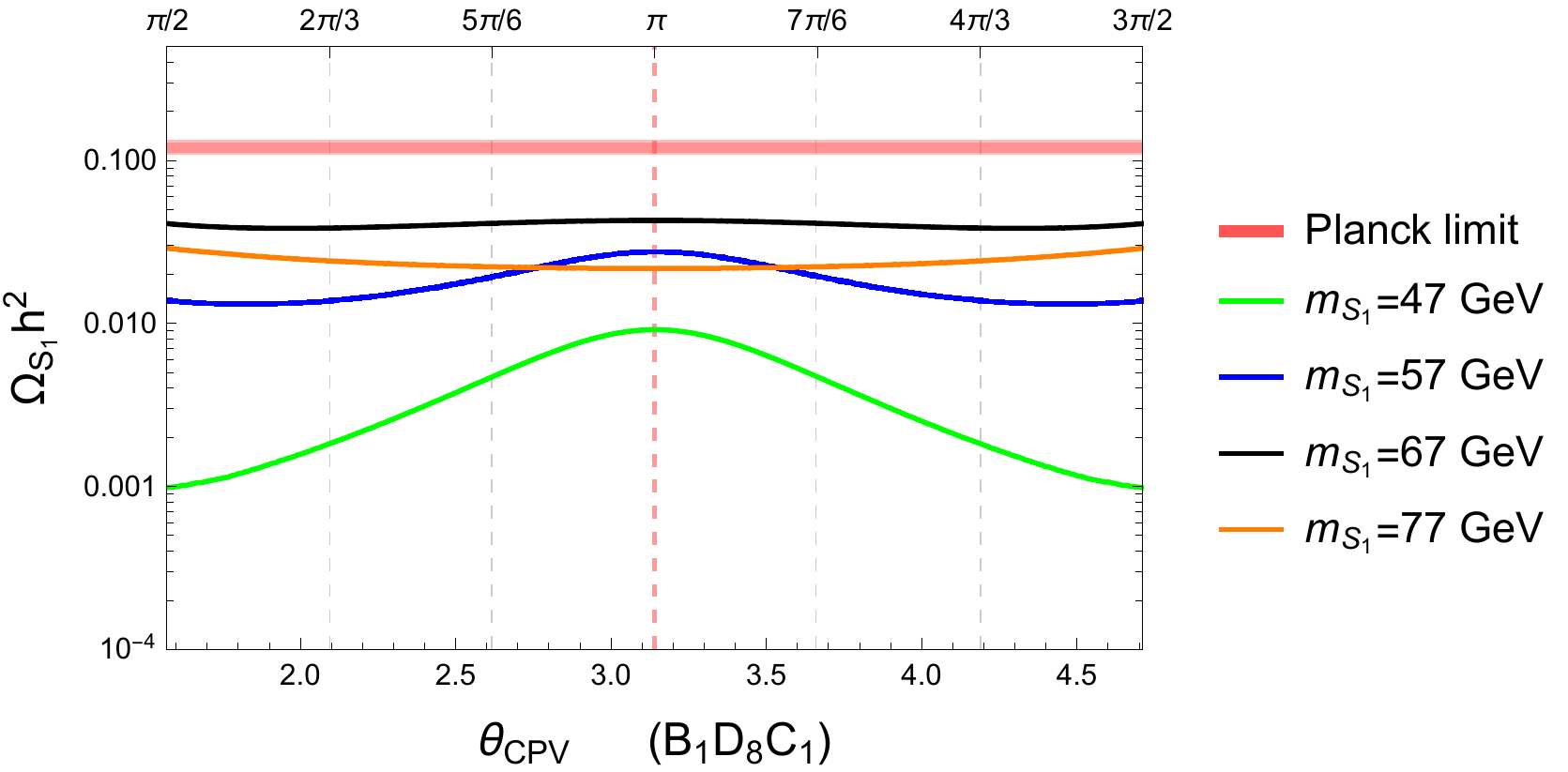}\\[2mm]
\includegraphics[scale=0.47]{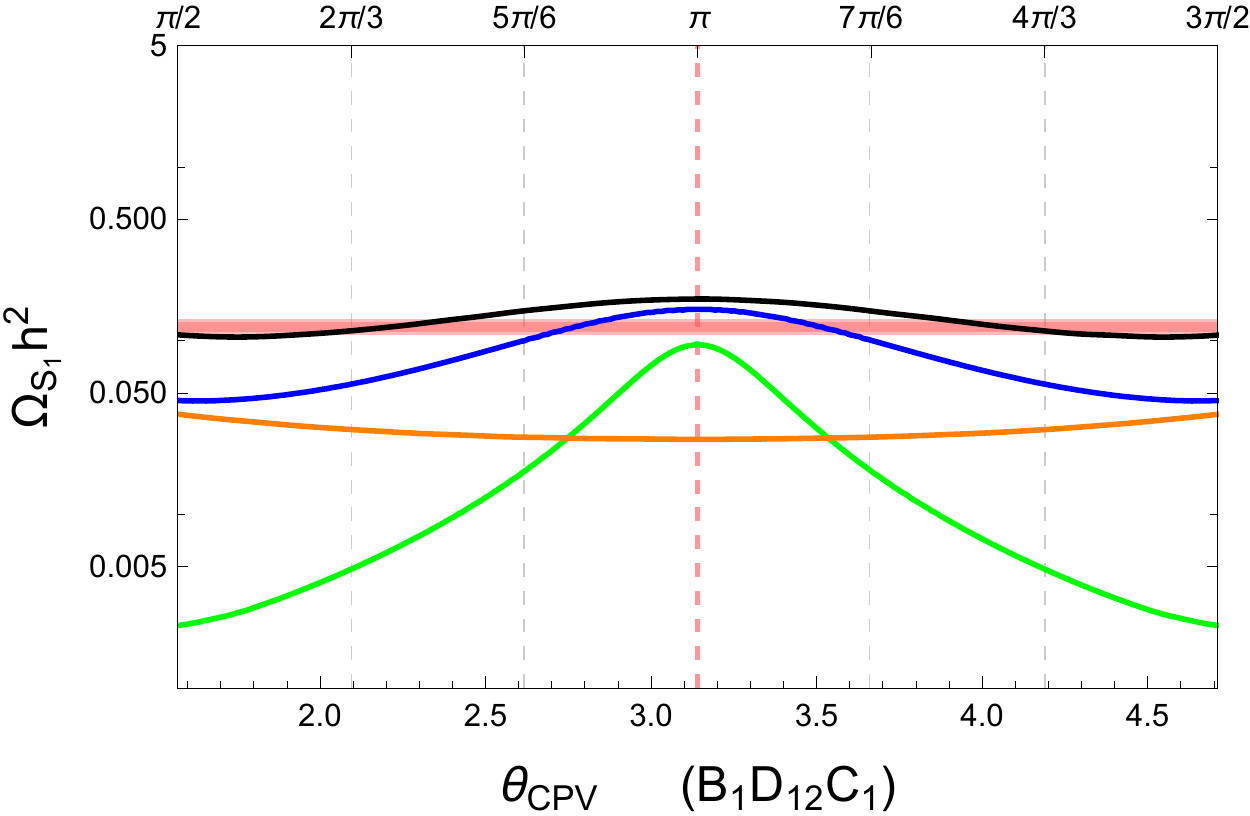}
\hspace{1mm}
\includegraphics[scale=0.47]{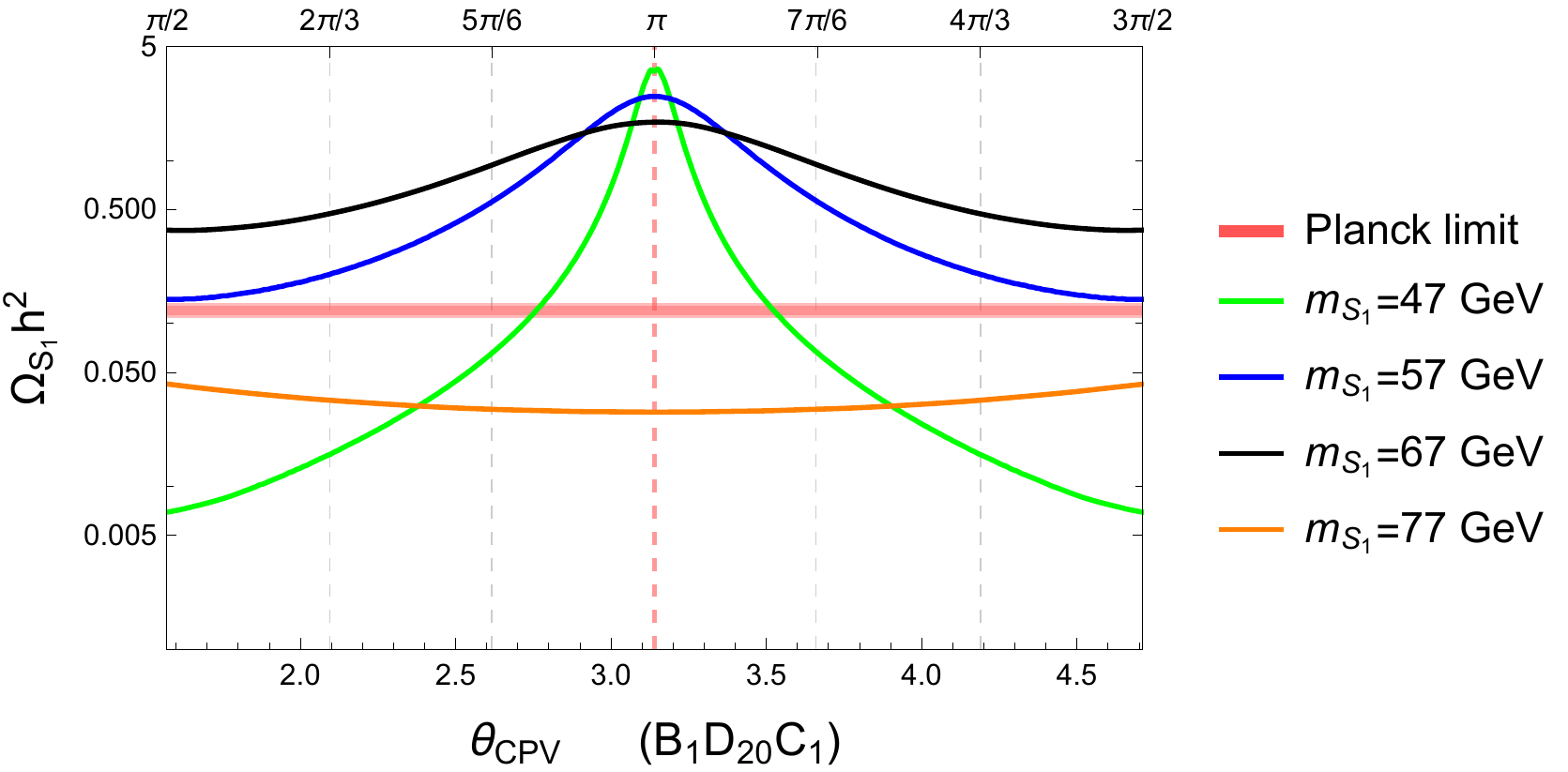}
\caption{The relic abundance of $S_1$ for different DM masses in type 1 benchmark scenarios. The horizontal red band shows the Planck observation limit on the abundance of DM.}
\label{B1-RelicAngle-fig}
\end{figure}

Note that as $\delta_{12}$ increases from 4 GeV in B$_1$D$_4$C$_1$ to 20 GeV in B$_1$D$_{20}$C$_1$, the co-annihilation probability of $S_1$ with other neutral dark particles is reduced, also the coupling of the main co-annihilation channel, $g_{ZS_1S_3}$ is reduced. As a result, the DM abundance is considerably larger in scenarios with larger $\delta_{12}$, which is well represented in figure \ref{B1-RelicAngle-fig}.

\begin{figure}[h!]
\begin{center}
\includegraphics[scale=0.46]{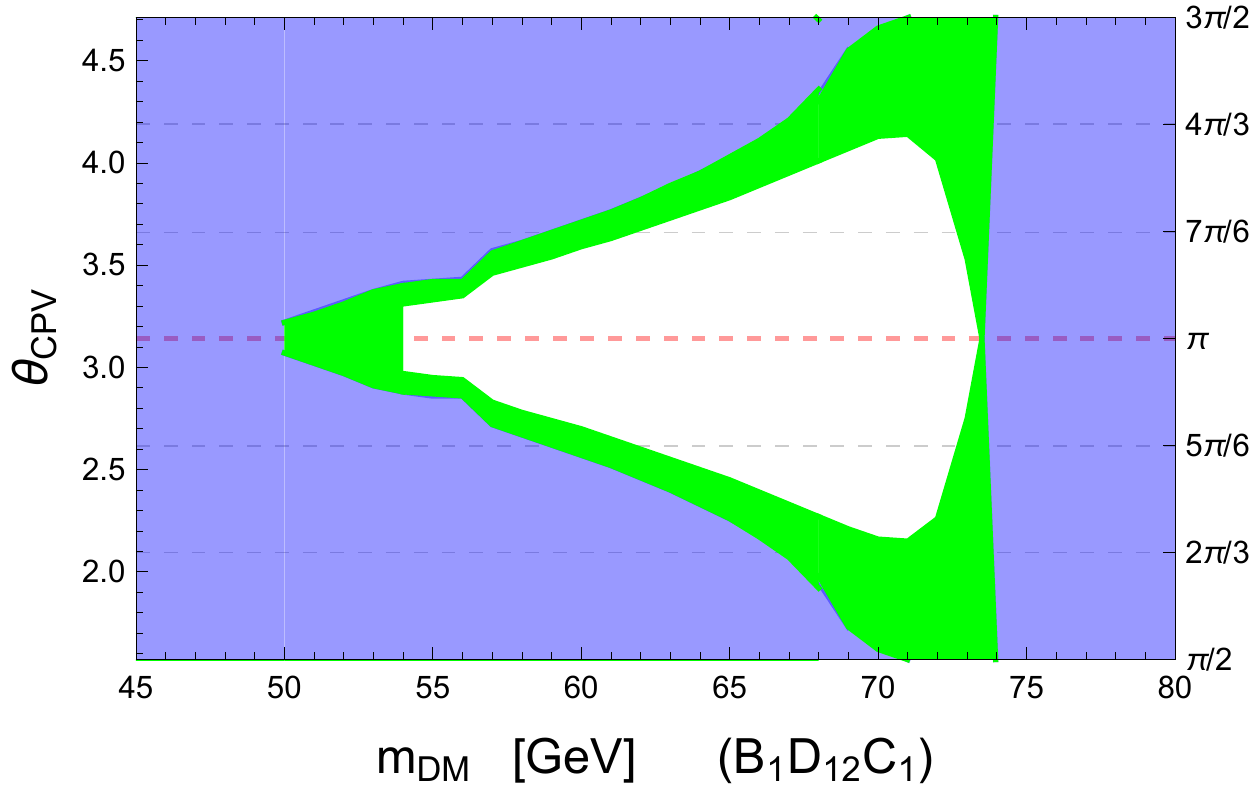}
\hspace{1mm}
\includegraphics[scale=0.46]{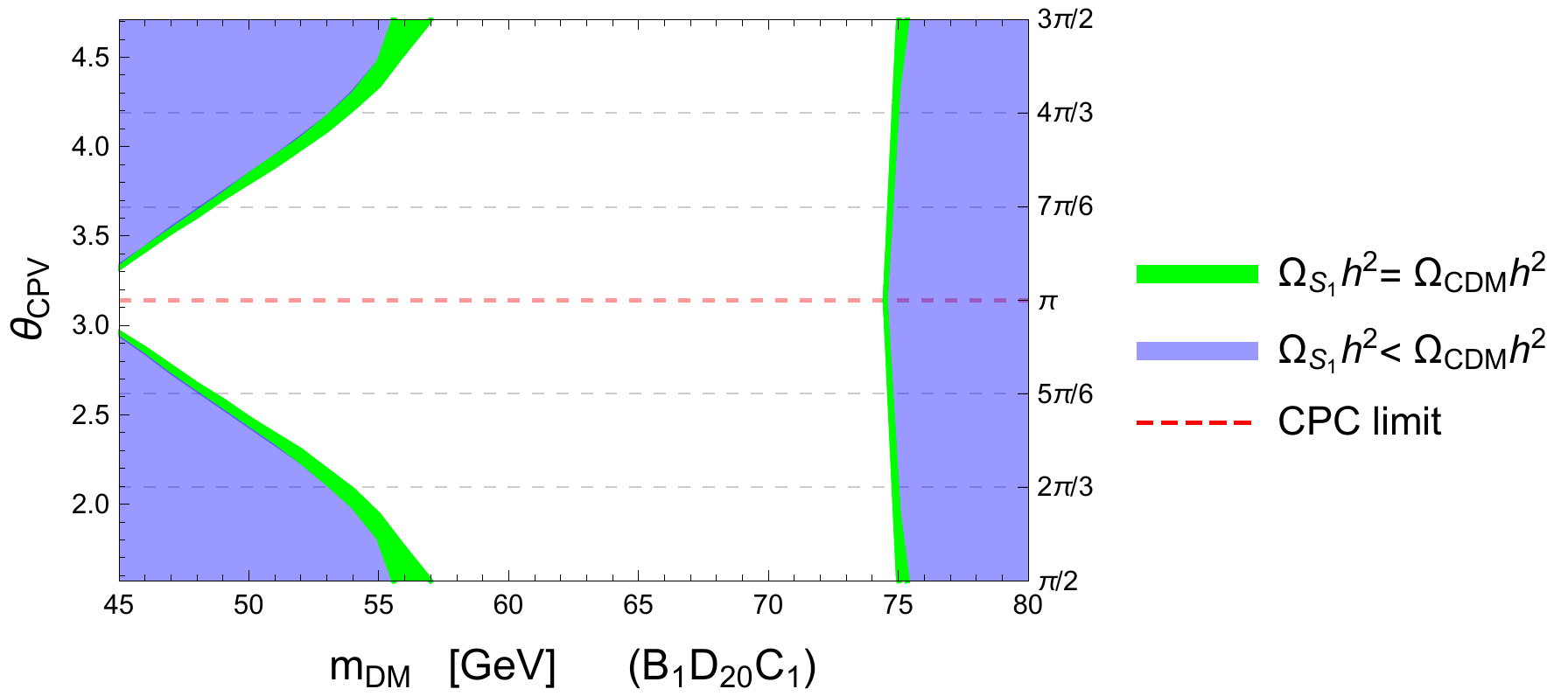}
\caption{Regions producing 100\% of DM in green, and under-producing DM in blue in the $\theta_{CPV}$-$m_{DM}$ plane in type 1 scenarios. The horizontal dashed red line represents the CP conserving limit.}
\label{B4D12-B4D20-AngleMass-fig}
\end{center}
\end{figure}

It is worth noting that in a given benchmark scenario as DM mass increases to values comparable with $m_{W^\pm}$ and $m_{Z}$, the $S_iS_i \to VV$ annihilation channels come into play and reduce the DM number density. 
As a result, DM is always under-produced for $m_{S_1} \gtrsim 80$ GeV. Figure \ref{B4D12-B4D20-AngleMass-fig} shows the regions with correct abundance and under-abundance of DM in B$_1$D$_{12}$C$_{1}$ and B$_1$D$_{20}$C$_{1}$ scenarios. In the other type 1 scenarios, B$_1$D$_{4}$C$_{1}$ and B$_1$D$_{8}$C$_{1}$, DM is always under-produced.
At $\theta_{CPV}=\pi$, where the model is reduced to the CP conserving limit, the value of $g_{S_iS_jZ}$ is fixed as a gauge coupling. With no handle on the $S_iS_jZ$ couplings, DM is over-produced in the intermediate mass region $54 \gev \lesssim m_{S_1} \lesssim 70 \gev $.
Varying the CP violating phase, $\theta_{CPV}$, can change the strength of the $S_iS_jZ$ couplings and, as a result, the $S_1$-$S_3$ mass splitting. The smallest $S_1$-$S_3$ mass splitting occurs close to the $\pi/2$ and $3\pi/2$ boundaries. Therefore, one expects a more effective $S_1$-$S_3$ co-annihilation and a smaller relic density as $\theta_{CPV}$ moves away from the CP conserving limit and towards the maximum CP violation at the $\pi/2$ and $3\pi/2$ boundaries.
Figure \ref{B4D12-B4D20-AngleMass-fig} illustrates this behaviour for B$_1$D$_{12}$C$_{1}$ and B$_1$D$_{20}$C$_{1}$ scenarios, where in the latter scenario with a large $\delta_{12}$, the intermediate mass region over-produces DM irrespective of the size of the $S_1$-$S_3$ mass splitting.

\subsection{Benchmarks of type 2}
\begin{figure}[h!]
\begin{center}
\includegraphics[scale=0.47]{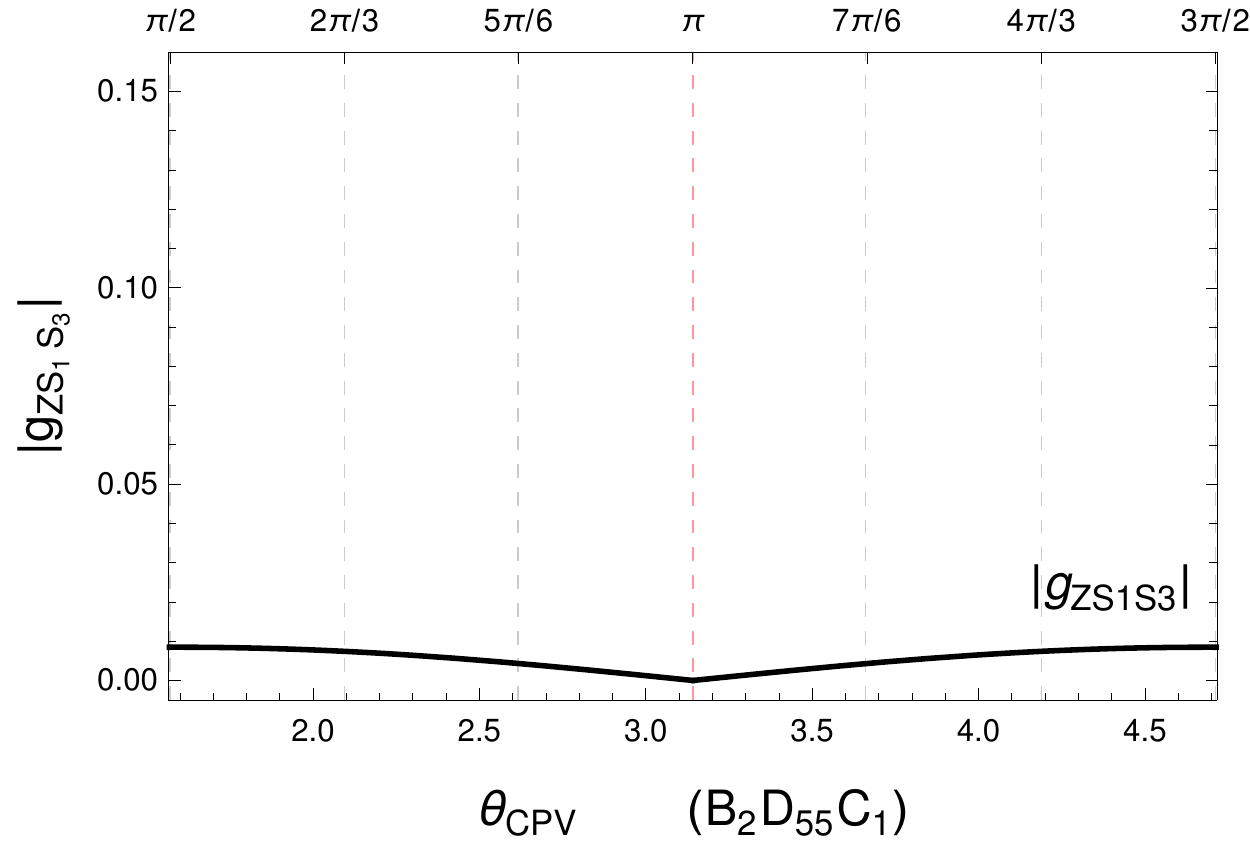}
\hspace{-2mm}
\includegraphics[scale=0.47]{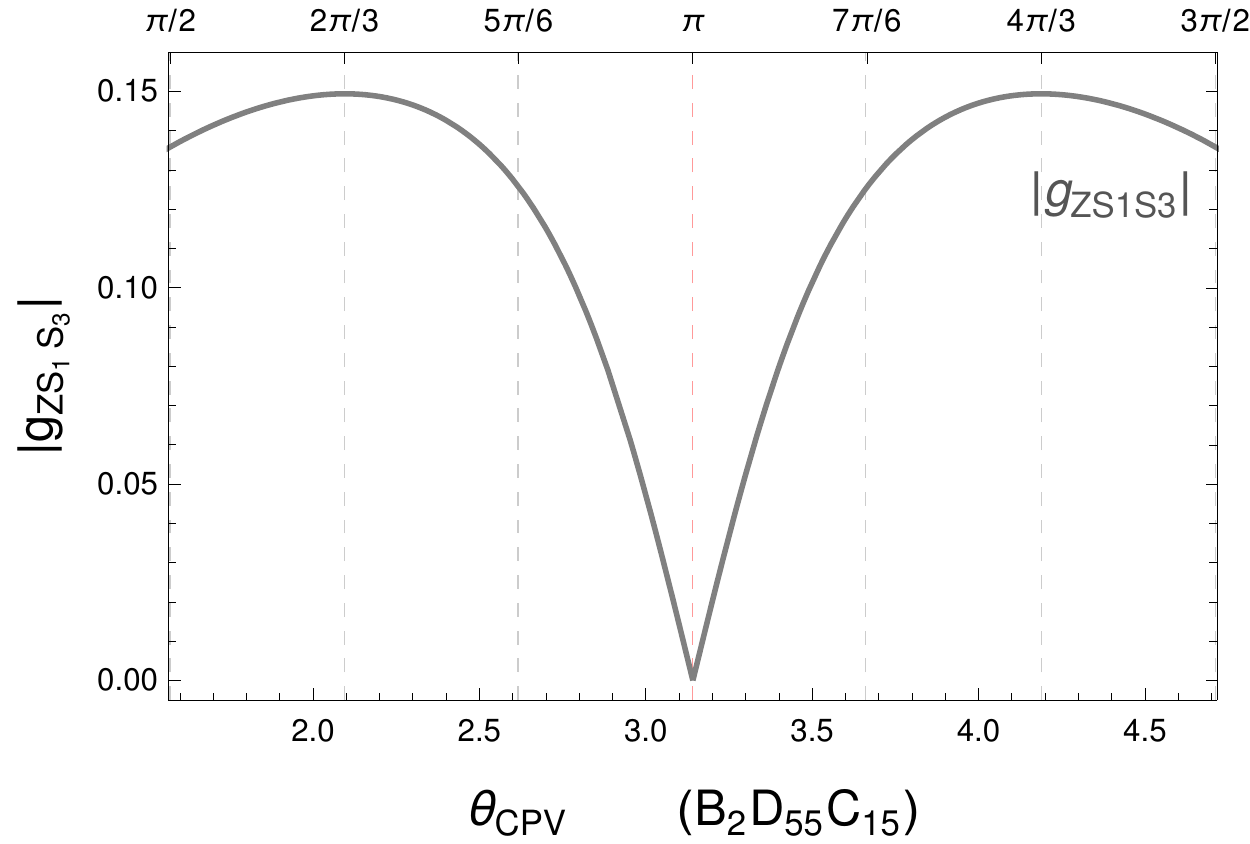}
\caption{The $g_{ZS_1S_3}-\theta_{CPV}$ dependence in type 2 benchmark scenarios for a given $m_{S_1}=50$ GeV.}
\label{B2D55-gZS1S3-fig}
\end{center}
\end{figure}

In type 2 benchmark scenarios,
B$_2$D$_{55}$C$_1$ and B$_2$D$_{55}$C$_{15}$, only $S_3$ is close in mass with $S_1$ and can co-annihilate with it through the $S_1S_3 \to Z^* \to f \bar f$ process, which dictates the behaviour of the model in the low mass region.
As the DM mass approaches the $ W^\pm,Z$ masses, the $S_1S_1 \to VV$ channels effectively annihilate the DM candidate leading to an insufficient relic density for $m_{DM}$ above this range, irrespective of the CP violating angle.
Recall that the $ZS_1S_3$ coupling is sensitive to the changes in the CP violating angle. Figure \ref{B2D55-gZS1S3-fig} shows the absolute value of the $ZS_1S_3$ coupling for an exemplary DM mass of 50 GeV with respect to $\theta_{CPV}$ in type 2 benchmark scenarios.

\begin{figure}[h!]
\begin{center}
\includegraphics[scale=0.47]{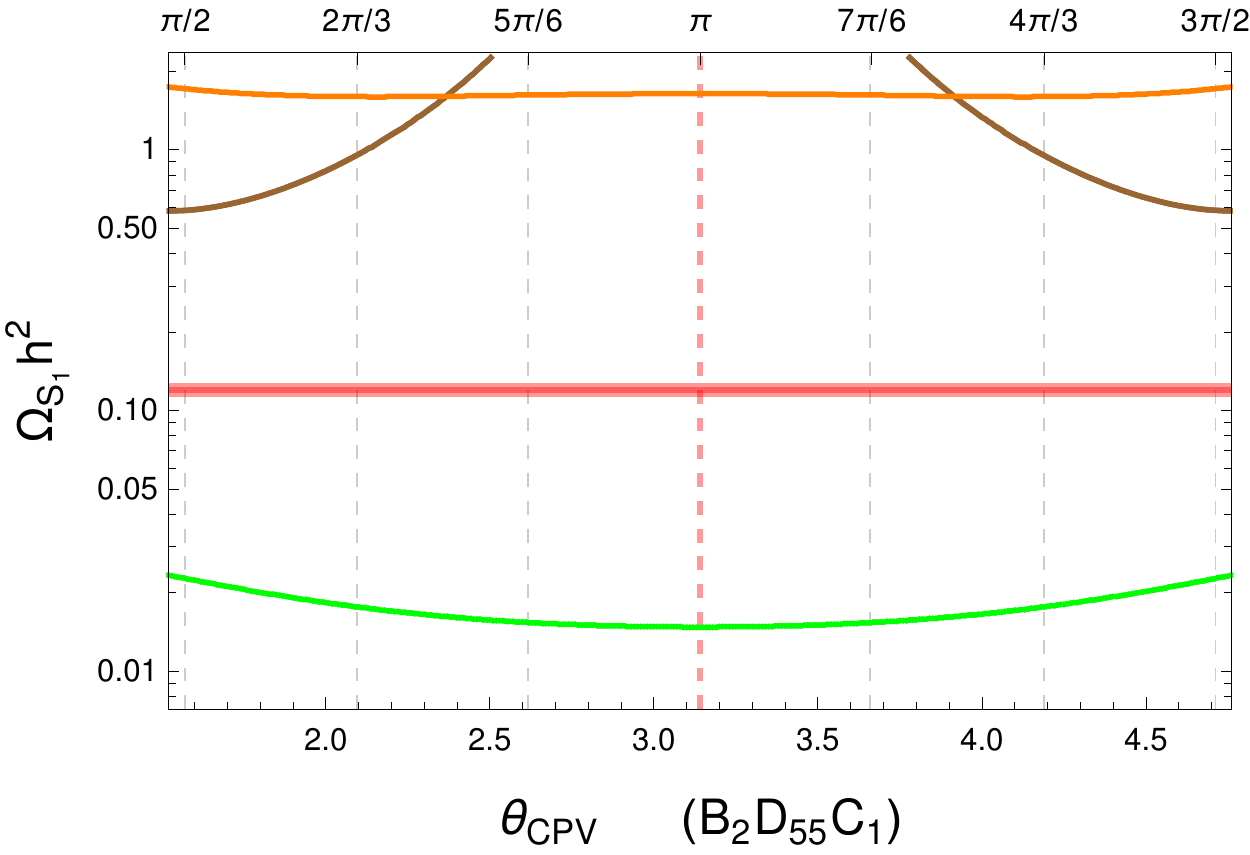}
\hspace{1mm}
\includegraphics[scale=0.47]{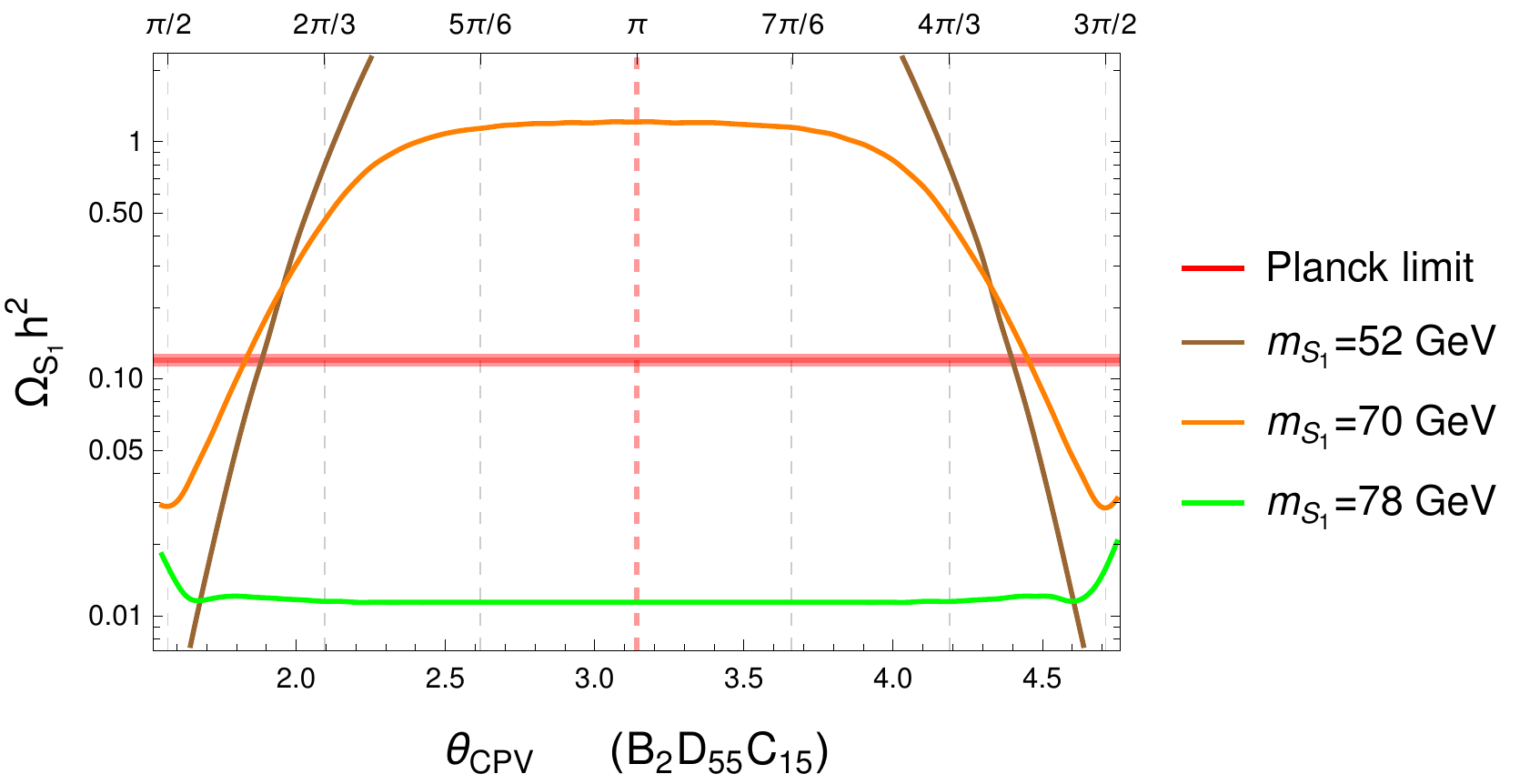}
\caption{The change in DM relic density for various DM masses with respect to the CP-violating angle. The horizontal red band shows the Planck observation limit on the abundance of DM.}
\label{B2D55-RelicAngle-fig}
\end{center}
\end{figure}

Due to the striking difference in the values of the $g_{ZS_1S_3}$ coupling in the two benchmark scenarios, one expects a substantial difference in the DM relic density in each case. 
In the mass range $45 \gev < m_{DM} < 75 \gev$, the B$_2$D$_{55}$C$_1$ scenario consistently over-produces DM  (except for large CP violating angles around the $Z$ resonance region $m_{DM} \approx m_Z/2$ where $S_1$ and $S_3$ are very close in mass) since the $ZS_1S_3$ coupling is so weak that it fails to co-annihilate DM effectively.
The B$_2$D$_{55}$C$_{15}$ scenario, on the other hand, has a large enough $ZS_1S_3$ coupling at large $\theta_{CPV}$ to satisfy the Planck limit on the DM relic density.
Figure \ref{B2D55-RelicAngle-fig} illustrates this behaviour where the relic density for various DM masses is shown. The B$_2$D$_{55}$C$_1$ scenario over-produces DM for masses below 75 GeV, while B$_2$D$_{55}$C$_{15}$ scenario produces DM in agreement with the Planck limit for large CP violating angles in this mass range. 
As mentioned before, both scenarios under-produce DM for larger masses when the $S_1S_1 \to VV$ annihilation channel is open.

\begin{figure}[h!]
\centering
\includegraphics[scale=0.46]{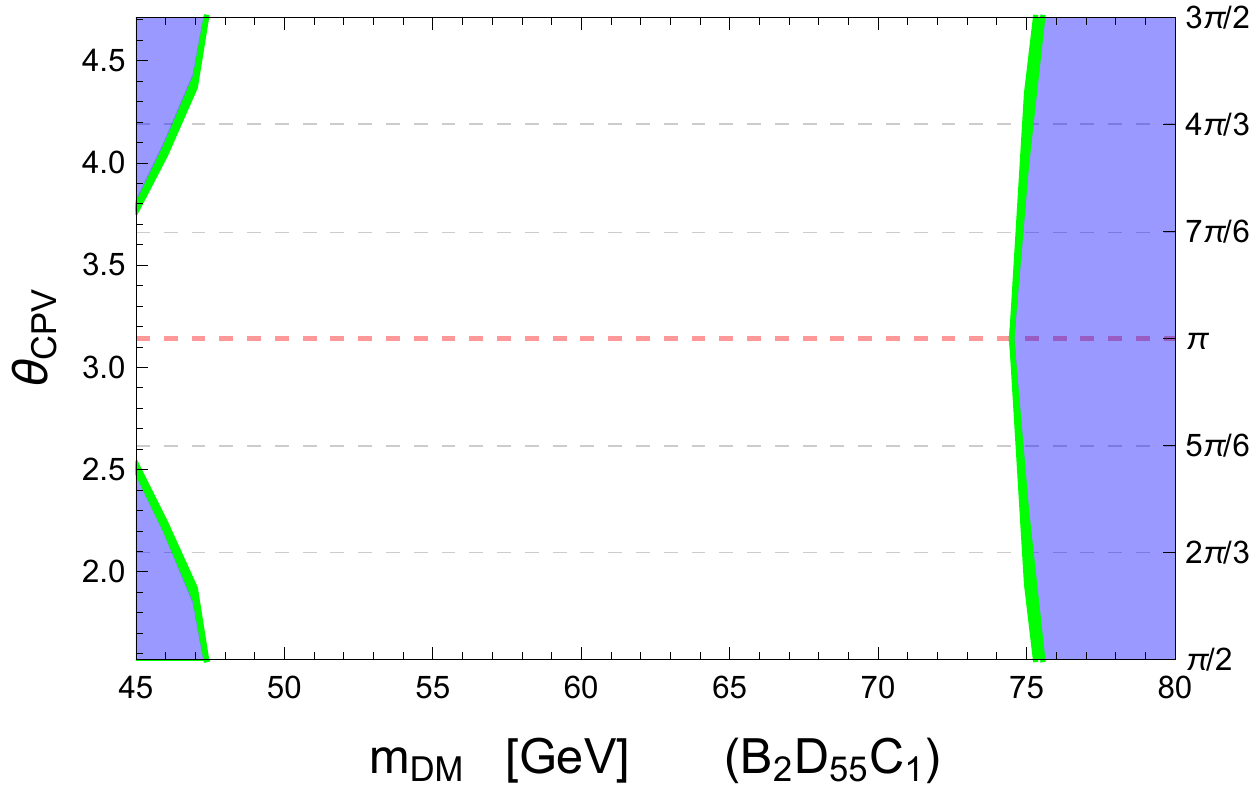}
\hspace{1mm}
\includegraphics[scale=0.46]{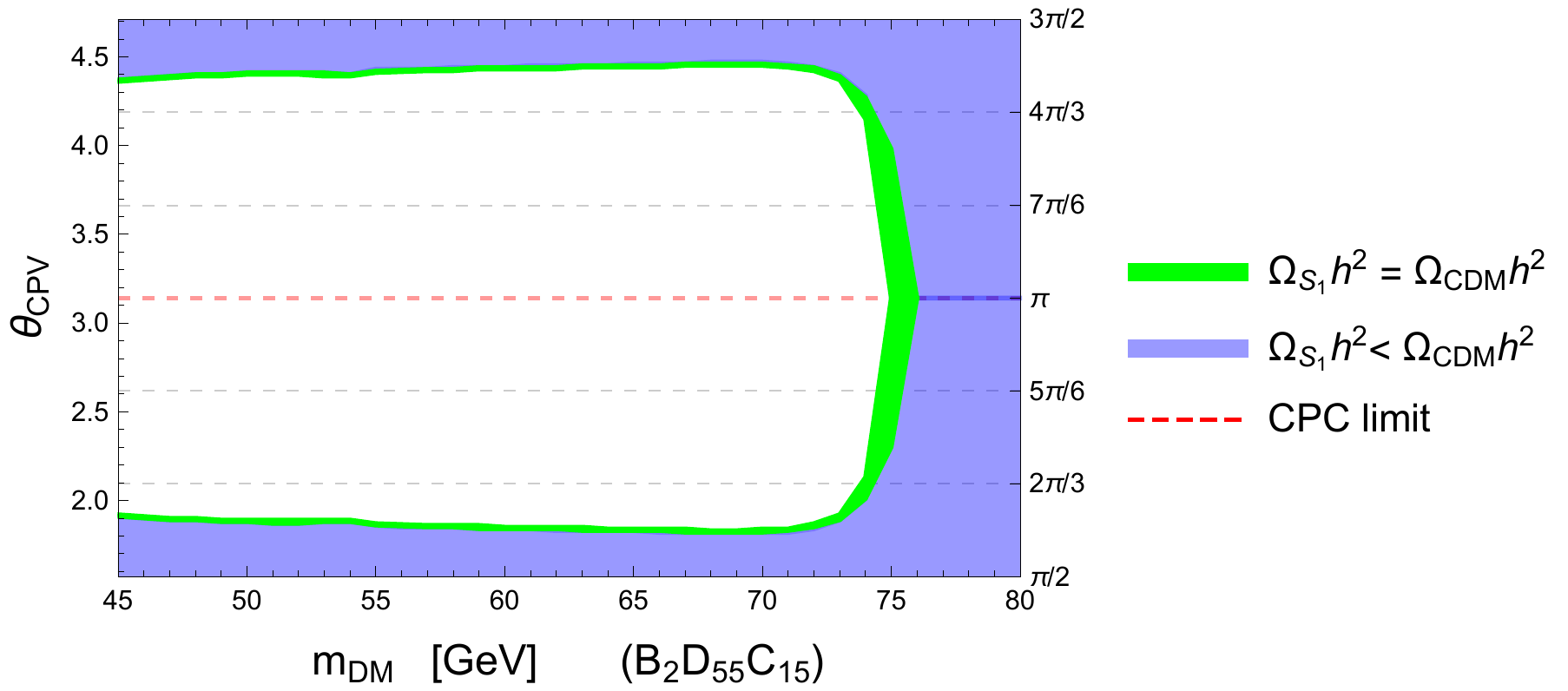}
\caption{Regions producing 100\% of DM in green, and under-producing DM in blue in the $\theta_{CPV}$-$m_{DM}$ plane in type 2 scenarios. The horizontal dashed red line represents the CP-conserving limit.}
\label{B2D55-AngleMass-fig}
\end{figure}

In figure \ref{B2D55-AngleMass-fig}, we show regions where $S_1$ contributes to 100\% of the observed DM in green, and regions where it only provides a fraction of the observed relic density in blue, in the $\theta_{CPV}$-$m_{DM}$ plane. The blank regions are ruled out by Planck observations as they lead to an over-production of DM. Note that in the CP conserving limit where $\theta_{CPV}=\pi$ the model fails to comply with the Planck observations throughout the low mass regime. 

\subsection{Benchmarks of type 3}
The type 3 benchmark scenarios which are defined in the heavy mass region, $m_{DM} \geqslant 80 \gev$, allow for $S_1$ to be close in mass with the charged inert scalars, thereby providing new co-annihilation channels for the DM candidate.
\begin{figure}[h!]
\begin{center}
\includegraphics[scale=0.47]{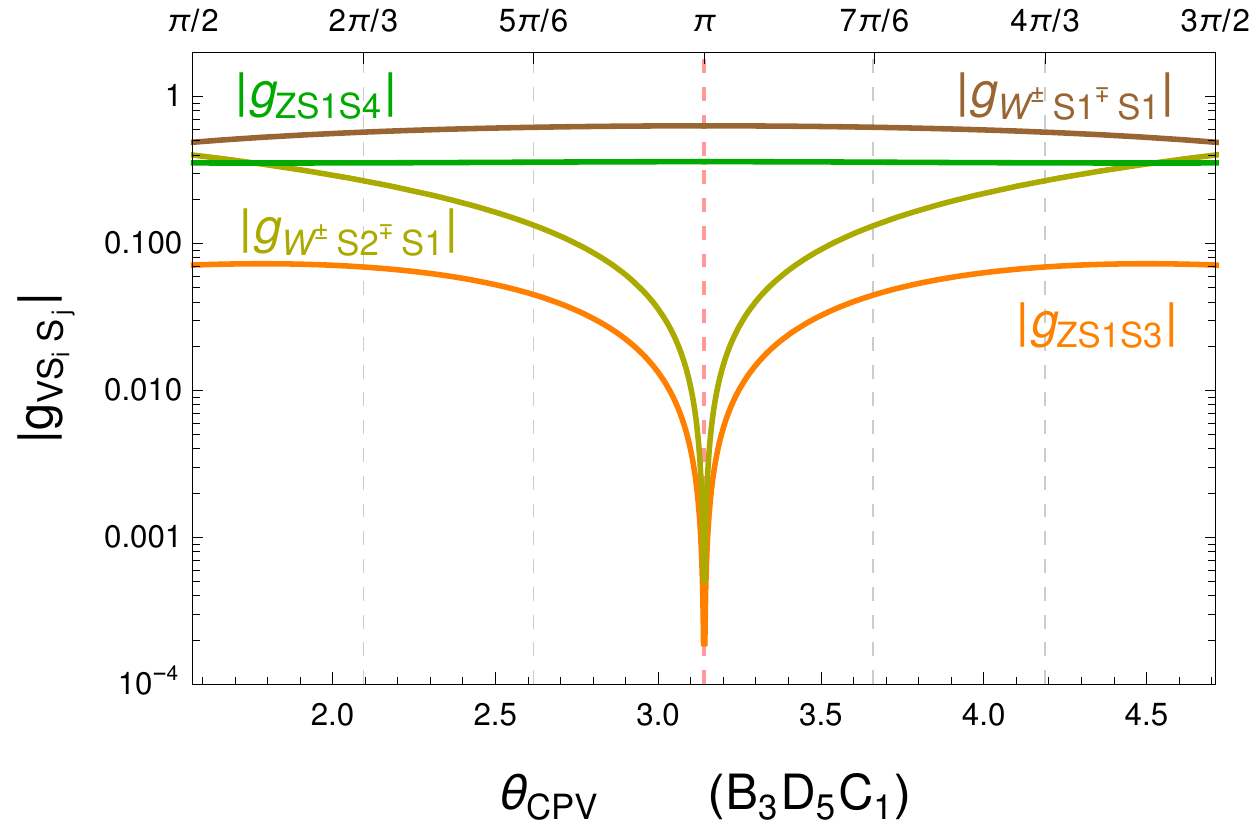}\\[2mm]
\includegraphics[scale=0.47]{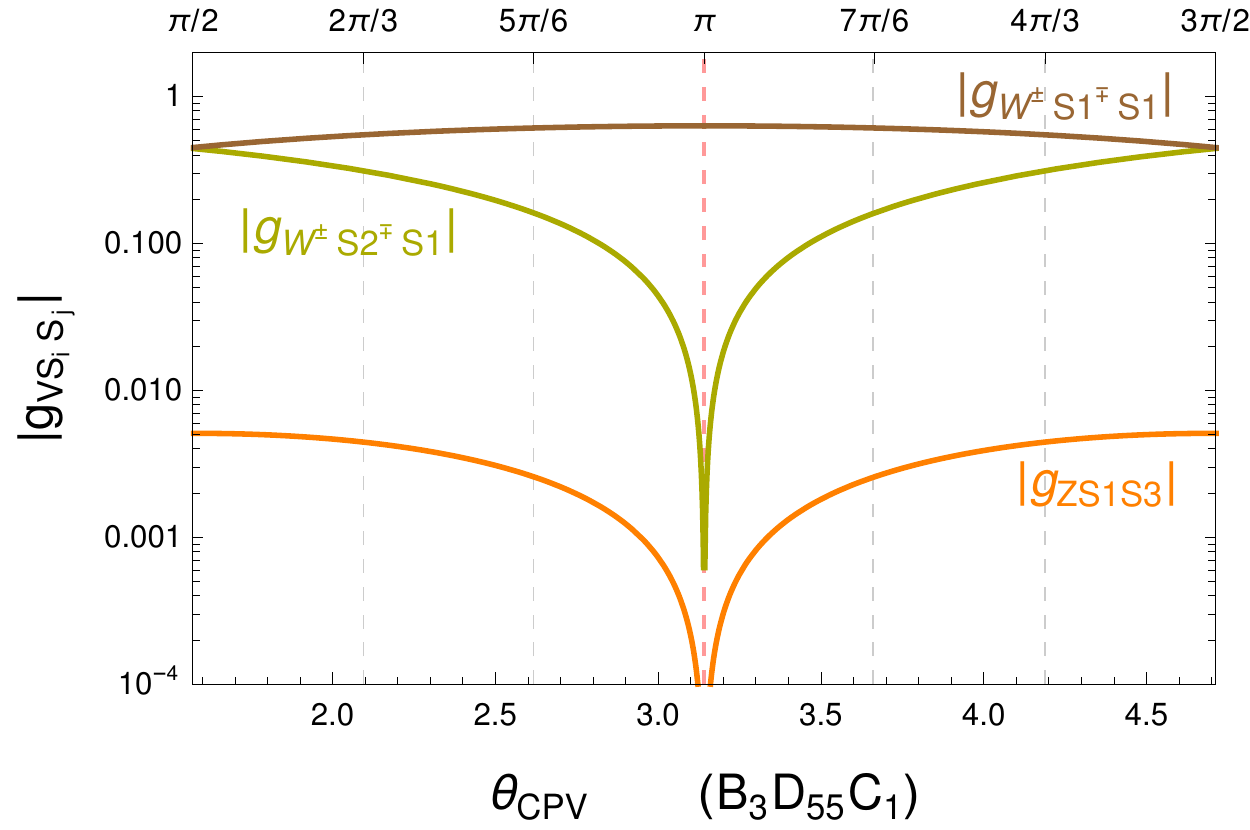}
\hspace{1mm}
\includegraphics[scale=0.47]{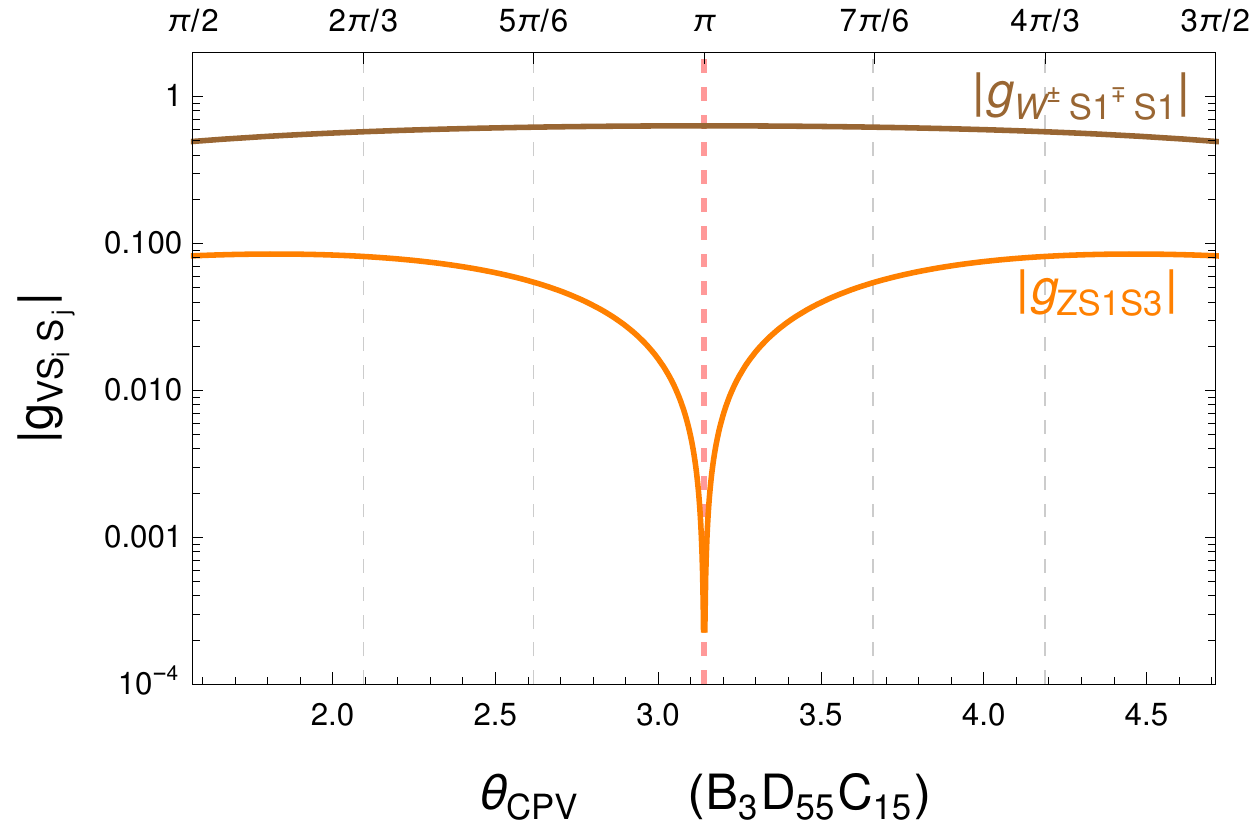}
\caption{The $g_{VS_iS_j}$-$\theta_{CPV}$ dependence in type 3 benchmark scenarios for a given $m_{S_1}=90$ GeV.}
\label{B3D55-gVSiSj-fig}
\end{center}
\end{figure}

When studying the DM phenomenology of the model, it is not only the annihilation and co-annihilation of DM, but also the (co)annihilation of other inert particles that should be taken into account.
Figure \ref{B3D55-gVSiSj-fig} shows the relevant and non-negligible $VS_1S_i$ couplings for all three type 3 benchmark scenarios where $V=W^\pm,Z$ and $S_{i}$ is a neutral or charged inert particle. Due to the presence of so many co-annihilation processes, type 3 scenarios consistently under-produce DM.
Moreover, in the heavy mass region the annihilation $S_1S_1 \to VV$ whose coupling is independent of the CP violating angle, is very effective and leads to the under-production of DM in all three scenarios.

\begin{figure}[h!]
\begin{center}
\includegraphics[scale=0.47]{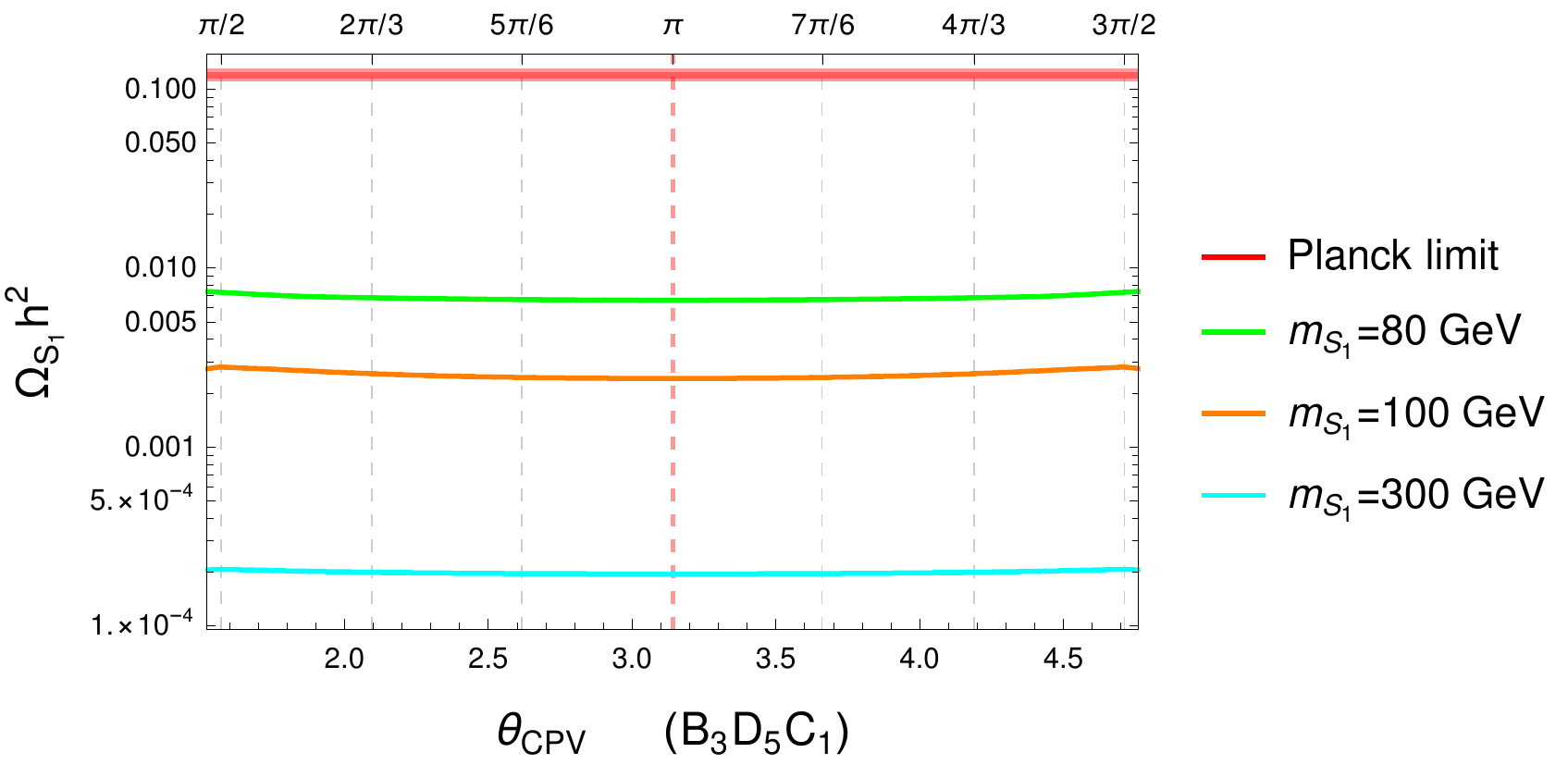}\\[2mm]
\includegraphics[scale=0.47]{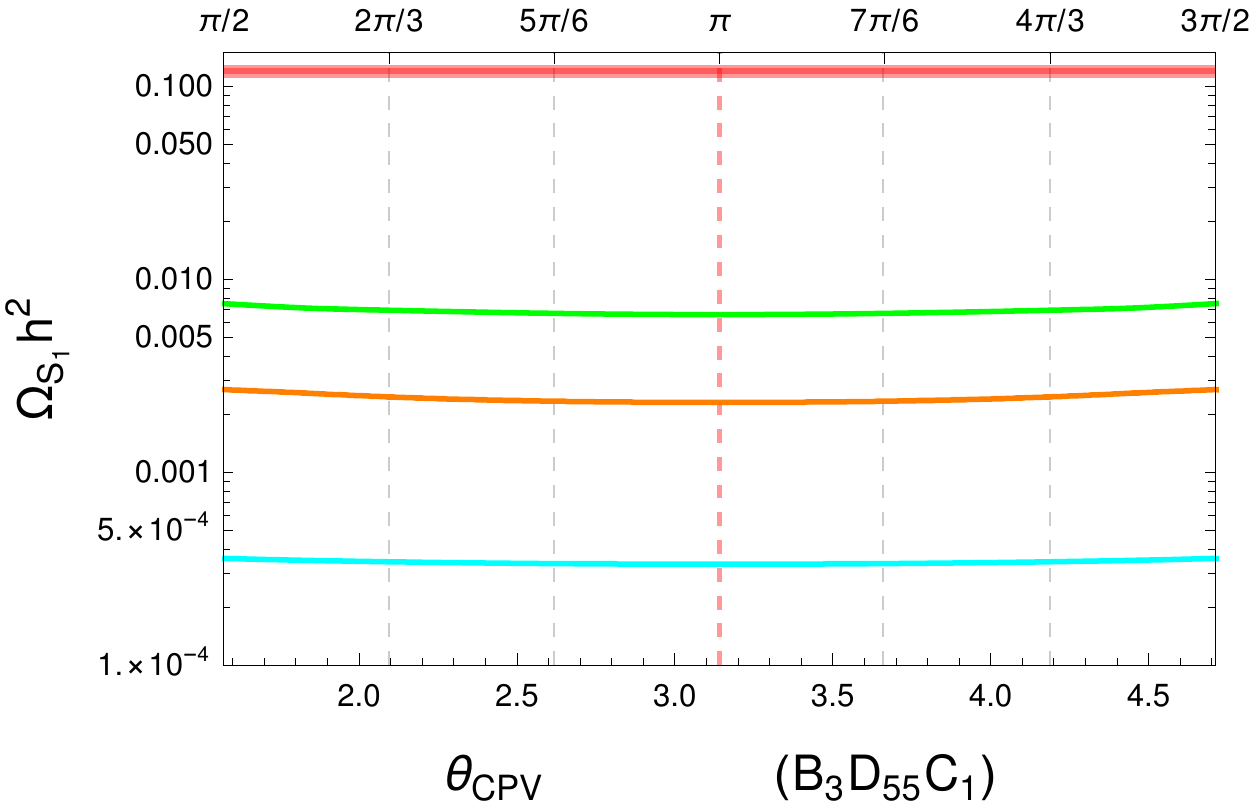}
\hspace{1mm}
\includegraphics[scale=0.47]{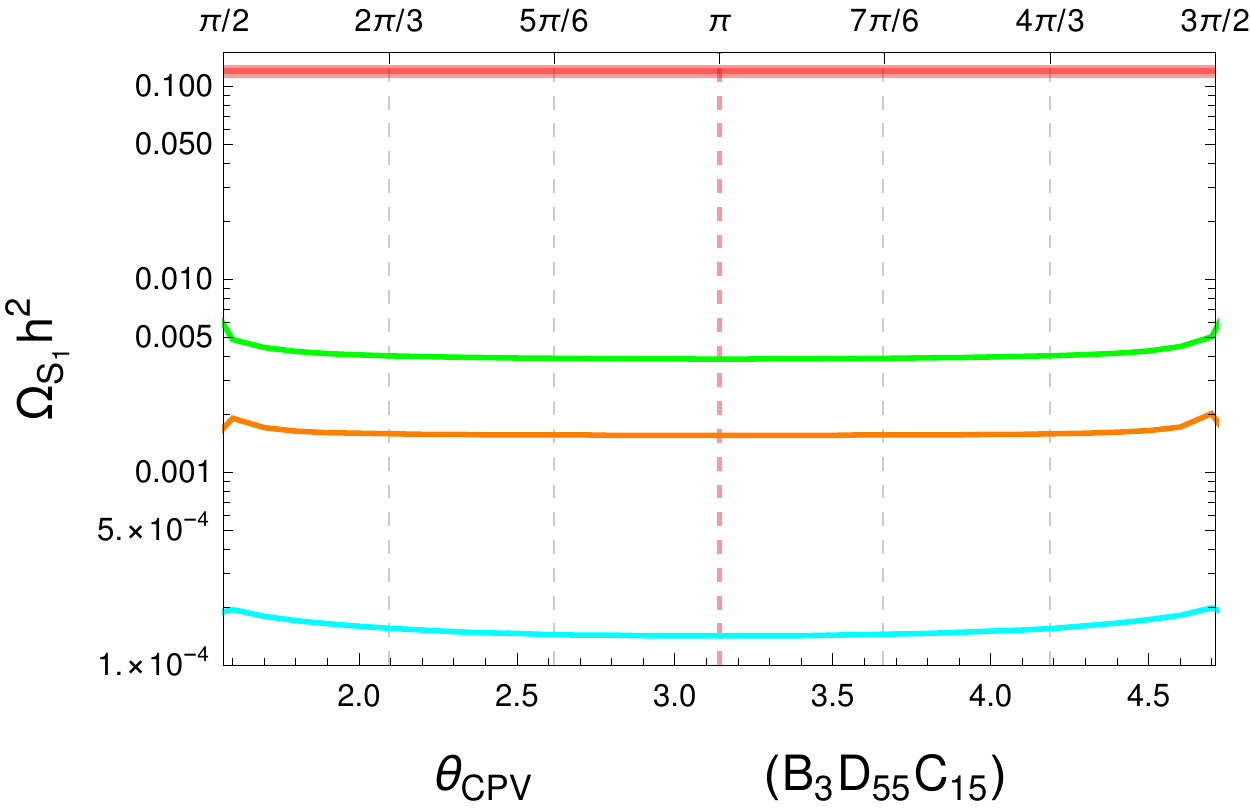}
\caption{The change DM relic density for various DM masses with respect to the CP-violating angle. The horizontal red band shows the Planck observation limit on the abundance of DM.}
\label{B3D5-B3D55-RelicAngle-fig}
\end{center}
\end{figure}

Figure \ref{B3D5-B3D55-RelicAngle-fig} illustrates the relic abundance for various DM masses. As mentioned before, all three scenarios have a similar behaviour, with the B$_3$D$_{55}$C$_1$ scenario providing a slightly larger relic density in comparison to the B$_3$D$_{5}$C$_1$ scenario due to a larger $\delta{12}$ mass splitting.
Scenario B$_3$D$_{55}$C$_{15}$ provides only two co-annihilation channels for $S_1$, however, they have larger couplings compared to the same processes in the B$_3$D$_{55}$C$_1$ case, which leads to a slightly smaller relic abundance for $S_1$ in the former scenario.
Let us point out that the reason we do not see the revival of the very heavy mass region $m_{DM} > 400$ GeV as shown in \cite{Keus:2015xya} and \cite{Cordero-Cid:2016krd} is the absence of the Higgs mediated processes 
$S_i S_j \to h \to VV$, where $S_{i,j}$ is any neutral or charged inert particle. These Higgs-mediated processes have a destructive interference with pure gauge processes $S_i S_j \to VV$, which would have reduced their (co)annihilation efficiency and revived the model in the heavy mass region.

\section{Conclusion and outlook}
\label{conclusion}

The scalar potential is the least constrained sector of the SM which, if extended, could provide viable DM candidates,  new sources of CP violation and a strong first order phase transition as the essential ingredients for EWBG. 
Of great importance are non-minimal Higgs frameworks with an extended dark/inert sector which could accommodate DM and dark CP violation unbounded by the EDM constraints, since the dark sector is protected by a conserved discrete symmetry from coupling to the SM fermions.

We study a well-motivated 3HDM with two inert and one active doublet to play the role of the SM Higgs doublet. The dark sector interacts with the visible sector through Higgs and SM gauge bosons. The couplings through Higgs are required to be small in agreement with direct and indirect detection experiments and SM-Higgs measurements, conversely, they need to be large enough for efficient (co)annihilation of DM.

We present a novel mechanism in which the CP violating dark particles only interact with the SM through the gauge bosons, primarily the $Z$ boson. Such $Z$-portal dark CP violation is realised in the regions of the parameter space where Higgs-mediated (co)annihilation processes are sub-dominant and have negligible contributions to the DM relic density. 
We show that in these regions of the parameter space, the $Z$ portal CP violating DM can still thermalise and satisfy all experimental and observational data. 

In the context of electroweak baryogenesis, the extended scalar sector could easily accommodate a strong first order phase transition. We discuss the efficient transfer of the unconstrained dark CP violation to the visible sector to source the matter-antimatter asymmetry in our upcoming publication.

\subsection*{Acknowledgement}
The author acknowledges financial support from Academy of Finland project ``Particle cosmology and gravitational waves" no.~320123 and would like to thank the organisers of the Corfu Summer Institute 2019 for the invitation and great hospitality.

\end{document}